\documentclass[12pt]{article}
\usepackage{epsfig,amsfonts,amssymb}
\usepackage{hyperref}
\pdfoutput=1

\usepackage{cite}
\usepackage{comment}

\def\ZZZ{{\hbox{ Z\kern-1.6mm Z}}}
\def\RRR{{\hbox{ R\kern-2.4mm R}}}
\def\CCC{{\hbox{ C\kern-2.0mm C}}}
\def\zzz{{\hbox{z\kern-1mm z}}}

\newcommand{\qeq}{{\hbox{=\kern-2.3mm ? \kern.5mm }}}
\renewcommand{\qeq}{=}

\newcommand{\eps}{\epsilon}

\newcommand{\II}{{\cal I}}

\newcommand{\OO}{{\cal O}}

\newcommand{\wt}{\widetilde}

\newcommand{\be}{\begin{equation}}
\newcommand{\ee}{\end{equation}}
\newcommand{\ben}{\begin{eqnarray}\displaystyle}
\newcommand{\een}{\end{eqnarray}}

\newcommand{\refb}[1]{(\ref{#1})}
\newcommand{\p}{\partial}
\newcommand{\sectiono}[1]{\section{#1}\setcounter{equation}{0}}

\def\one{{\hbox{ 1\kern-.8mm l}}}
\def\zero{{\hbox{ 0\kern-1.5mm 0}}}

\newcommand{\bea}[1]{\begin{eqnarray}\label{#1} }
\newcommand{\eea}{\end{eqnarray}}

\newcommand{\eqref}{\refb}

\usepackage{bm}
\usepackage[table]{xcolor}

\newcommand{\non}{\nonumber}

\def\figonea{
\def\JPicScale{0.4}
\ifx\JPicScale\undefined\def\JPicScale{1}\fi
\unitlength \JPicScale mm
\begin{picture}(135,90)(0,0)
\linethickness{0.3mm}
\multiput(45,90)(0.12,-0.18){167}{\line(0,-1){0.18}}
\linethickness{0.3mm}
\multiput(100,55)(0.12,0.12){250}{\line(1,0){0.12}}
\linethickness{0.3mm}
\multiput(100,25)(0.18,-0.12){167}{\line(1,0){0.18}}
\linethickness{0.3mm}
\multiput(30,5)(0.18,0.12){167}{\line(1,0){0.18}}

\put(80,40){\makebox(0,0)[cc]{$S$}}

\linethickness{0.3mm}
\put(105.56,40.03){\line(0,1){0.5}}
\multiput(105.55,41.03)(0.01,-0.5){1}{\line(0,-1){0.5}}
\multiput(105.53,41.53)(0.02,-0.5){1}{\line(0,-1){0.5}}
\multiput(105.5,42.04)(0.03,-0.5){1}{\line(0,-1){0.5}}
\multiput(105.46,42.54)(0.04,-0.5){1}{\line(0,-1){0.5}}
\multiput(105.41,43.04)(0.05,-0.5){1}{\line(0,-1){0.5}}
\multiput(105.35,43.54)(0.06,-0.5){1}{\line(0,-1){0.5}}
\multiput(105.28,44.03)(0.07,-0.5){1}{\line(0,-1){0.5}}
\multiput(105.2,44.53)(0.08,-0.5){1}{\line(0,-1){0.5}}
\multiput(105.11,45.02)(0.09,-0.49){1}{\line(0,-1){0.49}}
\multiput(105.01,45.52)(0.1,-0.49){1}{\line(0,-1){0.49}}
\multiput(104.9,46.01)(0.11,-0.49){1}{\line(0,-1){0.49}}
\multiput(104.78,46.5)(0.12,-0.49){1}{\line(0,-1){0.49}}
\multiput(104.65,46.98)(0.13,-0.49){1}{\line(0,-1){0.49}}
\multiput(104.51,47.47)(0.14,-0.48){1}{\line(0,-1){0.48}}
\multiput(104.37,47.95)(0.15,-0.48){1}{\line(0,-1){0.48}}
\multiput(104.21,48.42)(0.16,-0.48){1}{\line(0,-1){0.48}}
\multiput(104.04,48.9)(0.17,-0.47){1}{\line(0,-1){0.47}}
\multiput(103.87,49.37)(0.18,-0.47){1}{\line(0,-1){0.47}}
\multiput(103.68,49.84)(0.09,-0.23){2}{\line(0,-1){0.23}}
\multiput(103.49,50.3)(0.1,-0.23){2}{\line(0,-1){0.23}}
\multiput(103.28,50.76)(0.1,-0.23){2}{\line(0,-1){0.23}}
\multiput(103.07,51.21)(0.11,-0.23){2}{\line(0,-1){0.23}}
\multiput(102.85,51.66)(0.11,-0.23){2}{\line(0,-1){0.23}}
\multiput(102.62,52.11)(0.12,-0.22){2}{\line(0,-1){0.22}}
\multiput(102.38,52.55)(0.12,-0.22){2}{\line(0,-1){0.22}}
\multiput(102.13,52.99)(0.12,-0.22){2}{\line(0,-1){0.22}}
\multiput(101.87,53.42)(0.13,-0.22){2}{\line(0,-1){0.22}}
\multiput(101.61,53.85)(0.13,-0.21){2}{\line(0,-1){0.21}}
\multiput(101.33,54.27)(0.14,-0.21){2}{\line(0,-1){0.21}}
\multiput(101.05,54.69)(0.14,-0.21){2}{\line(0,-1){0.21}}
\multiput(100.76,55.1)(0.15,-0.21){2}{\line(0,-1){0.21}}
\multiput(100.46,55.5)(0.15,-0.2){2}{\line(0,-1){0.2}}
\multiput(100.15,55.9)(0.1,-0.13){3}{\line(0,-1){0.13}}
\multiput(99.84,56.29)(0.1,-0.13){3}{\line(0,-1){0.13}}
\multiput(99.52,56.68)(0.11,-0.13){3}{\line(0,-1){0.13}}
\multiput(99.19,57.06)(0.11,-0.13){3}{\line(0,-1){0.13}}
\multiput(98.85,57.43)(0.11,-0.12){3}{\line(0,-1){0.12}}
\multiput(98.5,57.79)(0.11,-0.12){3}{\line(0,-1){0.12}}
\multiput(98.15,58.15)(0.12,-0.12){3}{\line(0,-1){0.12}}
\multiput(97.79,58.5)(0.12,-0.12){3}{\line(1,0){0.12}}
\multiput(97.43,58.85)(0.12,-0.11){3}{\line(1,0){0.12}}
\multiput(97.06,59.19)(0.12,-0.11){3}{\line(1,0){0.12}}
\multiput(96.68,59.52)(0.13,-0.11){3}{\line(1,0){0.13}}
\multiput(96.29,59.84)(0.13,-0.11){3}{\line(1,0){0.13}}
\multiput(95.9,60.15)(0.13,-0.1){3}{\line(1,0){0.13}}
\multiput(95.5,60.46)(0.13,-0.1){3}{\line(1,0){0.13}}
\multiput(95.1,60.76)(0.2,-0.15){2}{\line(1,0){0.2}}
\multiput(94.69,61.05)(0.21,-0.15){2}{\line(1,0){0.21}}
\multiput(94.27,61.33)(0.21,-0.14){2}{\line(1,0){0.21}}
\multiput(93.85,61.61)(0.21,-0.14){2}{\line(1,0){0.21}}
\multiput(93.42,61.87)(0.21,-0.13){2}{\line(1,0){0.21}}
\multiput(92.99,62.13)(0.22,-0.13){2}{\line(1,0){0.22}}
\multiput(92.55,62.38)(0.22,-0.12){2}{\line(1,0){0.22}}
\multiput(92.11,62.62)(0.22,-0.12){2}{\line(1,0){0.22}}
\multiput(91.66,62.85)(0.22,-0.12){2}{\line(1,0){0.22}}
\multiput(91.21,63.07)(0.23,-0.11){2}{\line(1,0){0.23}}
\multiput(90.76,63.28)(0.23,-0.11){2}{\line(1,0){0.23}}
\multiput(90.3,63.49)(0.23,-0.1){2}{\line(1,0){0.23}}
\multiput(89.84,63.68)(0.23,-0.1){2}{\line(1,0){0.23}}
\multiput(89.37,63.87)(0.23,-0.09){2}{\line(1,0){0.23}}
\multiput(88.9,64.04)(0.47,-0.18){1}{\line(1,0){0.47}}
\multiput(88.42,64.21)(0.47,-0.17){1}{\line(1,0){0.47}}
\multiput(87.95,64.37)(0.48,-0.16){1}{\line(1,0){0.48}}
\multiput(87.47,64.51)(0.48,-0.15){1}{\line(1,0){0.48}}
\multiput(86.98,64.65)(0.48,-0.14){1}{\line(1,0){0.48}}
\multiput(86.5,64.78)(0.49,-0.13){1}{\line(1,0){0.49}}
\multiput(86.01,64.9)(0.49,-0.12){1}{\line(1,0){0.49}}
\multiput(85.52,65.01)(0.49,-0.11){1}{\line(1,0){0.49}}
\multiput(85.02,65.11)(0.49,-0.1){1}{\line(1,0){0.49}}
\multiput(84.53,65.2)(0.49,-0.09){1}{\line(1,0){0.49}}
\multiput(84.03,65.28)(0.5,-0.08){1}{\line(1,0){0.5}}
\multiput(83.54,65.35)(0.5,-0.07){1}{\line(1,0){0.5}}
\multiput(83.04,65.41)(0.5,-0.06){1}{\line(1,0){0.5}}
\multiput(82.54,65.46)(0.5,-0.05){1}{\line(1,0){0.5}}
\multiput(82.04,65.5)(0.5,-0.04){1}{\line(1,0){0.5}}
\multiput(81.53,65.53)(0.5,-0.03){1}{\line(1,0){0.5}}
\multiput(81.03,65.55)(0.5,-0.02){1}{\line(1,0){0.5}}
\multiput(80.53,65.56)(0.5,-0.01){1}{\line(1,0){0.5}}
\put(80.03,65.56){\line(1,0){0.5}}
\multiput(79.52,65.55)(0.5,0.01){1}{\line(1,0){0.5}}
\multiput(79.02,65.53)(0.5,0.02){1}{\line(1,0){0.5}}
\multiput(78.52,65.5)(0.5,0.03){1}{\line(1,0){0.5}}
\multiput(78.02,65.46)(0.5,0.04){1}{\line(1,0){0.5}}
\multiput(77.52,65.41)(0.5,0.05){1}{\line(1,0){0.5}}
\multiput(77.02,65.35)(0.5,0.06){1}{\line(1,0){0.5}}
\multiput(76.52,65.28)(0.5,0.07){1}{\line(1,0){0.5}}
\multiput(76.03,65.2)(0.5,0.08){1}{\line(1,0){0.5}}
\multiput(75.53,65.11)(0.49,0.09){1}{\line(1,0){0.49}}
\multiput(75.04,65.01)(0.49,0.1){1}{\line(1,0){0.49}}
\multiput(74.55,64.9)(0.49,0.11){1}{\line(1,0){0.49}}
\multiput(74.06,64.78)(0.49,0.12){1}{\line(1,0){0.49}}
\multiput(73.57,64.65)(0.49,0.13){1}{\line(1,0){0.49}}
\multiput(73.09,64.51)(0.48,0.14){1}{\line(1,0){0.48}}
\multiput(72.61,64.37)(0.48,0.15){1}{\line(1,0){0.48}}
\multiput(72.13,64.21)(0.48,0.16){1}{\line(1,0){0.48}}
\multiput(71.66,64.04)(0.47,0.17){1}{\line(1,0){0.47}}
\multiput(71.19,63.87)(0.47,0.18){1}{\line(1,0){0.47}}
\multiput(70.72,63.68)(0.23,0.09){2}{\line(1,0){0.23}}
\multiput(70.26,63.49)(0.23,0.1){2}{\line(1,0){0.23}}
\multiput(69.8,63.28)(0.23,0.1){2}{\line(1,0){0.23}}
\multiput(69.34,63.07)(0.23,0.11){2}{\line(1,0){0.23}}
\multiput(68.89,62.85)(0.23,0.11){2}{\line(1,0){0.23}}
\multiput(68.44,62.62)(0.22,0.12){2}{\line(1,0){0.22}}
\multiput(68,62.38)(0.22,0.12){2}{\line(1,0){0.22}}
\multiput(67.57,62.13)(0.22,0.12){2}{\line(1,0){0.22}}
\multiput(67.13,61.87)(0.22,0.13){2}{\line(1,0){0.22}}
\multiput(66.71,61.61)(0.21,0.13){2}{\line(1,0){0.21}}
\multiput(66.29,61.33)(0.21,0.14){2}{\line(1,0){0.21}}
\multiput(65.87,61.05)(0.21,0.14){2}{\line(1,0){0.21}}
\multiput(65.46,60.76)(0.21,0.15){2}{\line(1,0){0.21}}
\multiput(65.06,60.46)(0.2,0.15){2}{\line(1,0){0.2}}
\multiput(64.66,60.15)(0.13,0.1){3}{\line(1,0){0.13}}
\multiput(64.27,59.84)(0.13,0.1){3}{\line(1,0){0.13}}
\multiput(63.88,59.52)(0.13,0.11){3}{\line(1,0){0.13}}
\multiput(63.5,59.19)(0.13,0.11){3}{\line(1,0){0.13}}
\multiput(63.13,58.85)(0.12,0.11){3}{\line(1,0){0.12}}
\multiput(62.76,58.5)(0.12,0.11){3}{\line(1,0){0.12}}
\multiput(62.4,58.15)(0.12,0.12){3}{\line(1,0){0.12}}
\multiput(62.05,57.79)(0.12,0.12){3}{\line(0,1){0.12}}
\multiput(61.71,57.43)(0.11,0.12){3}{\line(0,1){0.12}}
\multiput(61.37,57.06)(0.11,0.12){3}{\line(0,1){0.12}}
\multiput(61.04,56.68)(0.11,0.13){3}{\line(0,1){0.13}}
\multiput(60.72,56.29)(0.11,0.13){3}{\line(0,1){0.13}}
\multiput(60.4,55.9)(0.1,0.13){3}{\line(0,1){0.13}}
\multiput(60.1,55.5)(0.1,0.13){3}{\line(0,1){0.13}}
\multiput(59.8,55.1)(0.15,0.2){2}{\line(0,1){0.2}}
\multiput(59.51,54.69)(0.15,0.21){2}{\line(0,1){0.21}}
\multiput(59.22,54.27)(0.14,0.21){2}{\line(0,1){0.21}}
\multiput(58.95,53.85)(0.14,0.21){2}{\line(0,1){0.21}}
\multiput(58.68,53.42)(0.13,0.21){2}{\line(0,1){0.21}}
\multiput(58.43,52.99)(0.13,0.22){2}{\line(0,1){0.22}}
\multiput(58.18,52.55)(0.12,0.22){2}{\line(0,1){0.22}}
\multiput(57.94,52.11)(0.12,0.22){2}{\line(0,1){0.22}}
\multiput(57.71,51.66)(0.12,0.22){2}{\line(0,1){0.22}}
\multiput(57.49,51.21)(0.11,0.23){2}{\line(0,1){0.23}}
\multiput(57.27,50.76)(0.11,0.23){2}{\line(0,1){0.23}}
\multiput(57.07,50.3)(0.1,0.23){2}{\line(0,1){0.23}}
\multiput(56.87,49.84)(0.1,0.23){2}{\line(0,1){0.23}}
\multiput(56.69,49.37)(0.09,0.23){2}{\line(0,1){0.23}}
\multiput(56.51,48.9)(0.18,0.47){1}{\line(0,1){0.47}}
\multiput(56.35,48.42)(0.17,0.47){1}{\line(0,1){0.47}}
\multiput(56.19,47.95)(0.16,0.48){1}{\line(0,1){0.48}}
\multiput(56.04,47.47)(0.15,0.48){1}{\line(0,1){0.48}}
\multiput(55.9,46.98)(0.14,0.48){1}{\line(0,1){0.48}}
\multiput(55.78,46.5)(0.13,0.49){1}{\line(0,1){0.49}}
\multiput(55.66,46.01)(0.12,0.49){1}{\line(0,1){0.49}}
\multiput(55.55,45.52)(0.11,0.49){1}{\line(0,1){0.49}}
\multiput(55.45,45.02)(0.1,0.49){1}{\line(0,1){0.49}}
\multiput(55.36,44.53)(0.09,0.49){1}{\line(0,1){0.49}}
\multiput(55.28,44.03)(0.08,0.5){1}{\line(0,1){0.5}}
\multiput(55.21,43.54)(0.07,0.5){1}{\line(0,1){0.5}}
\multiput(55.15,43.04)(0.06,0.5){1}{\line(0,1){0.5}}
\multiput(55.1,42.54)(0.05,0.5){1}{\line(0,1){0.5}}
\multiput(55.06,42.04)(0.04,0.5){1}{\line(0,1){0.5}}
\multiput(55.03,41.53)(0.03,0.5){1}{\line(0,1){0.5}}
\multiput(55.01,41.03)(0.02,0.5){1}{\line(0,1){0.5}}
\multiput(55,40.53)(0.01,0.5){1}{\line(0,1){0.5}}
\put(55,40.03){\line(0,1){0.5}}
\multiput(55,40.03)(0.01,-0.5){1}{\line(0,-1){0.5}}
\multiput(55.01,39.52)(0.02,-0.5){1}{\line(0,-1){0.5}}
\multiput(55.03,39.02)(0.03,-0.5){1}{\line(0,-1){0.5}}
\multiput(55.06,38.52)(0.04,-0.5){1}{\line(0,-1){0.5}}
\multiput(55.1,38.02)(0.05,-0.5){1}{\line(0,-1){0.5}}
\multiput(55.15,37.52)(0.06,-0.5){1}{\line(0,-1){0.5}}
\multiput(55.21,37.02)(0.07,-0.5){1}{\line(0,-1){0.5}}
\multiput(55.28,36.52)(0.08,-0.5){1}{\line(0,-1){0.5}}
\multiput(55.36,36.03)(0.09,-0.49){1}{\line(0,-1){0.49}}
\multiput(55.45,35.53)(0.1,-0.49){1}{\line(0,-1){0.49}}
\multiput(55.55,35.04)(0.11,-0.49){1}{\line(0,-1){0.49}}
\multiput(55.66,34.55)(0.12,-0.49){1}{\line(0,-1){0.49}}
\multiput(55.78,34.06)(0.13,-0.49){1}{\line(0,-1){0.49}}
\multiput(55.9,33.57)(0.14,-0.48){1}{\line(0,-1){0.48}}
\multiput(56.04,33.09)(0.15,-0.48){1}{\line(0,-1){0.48}}
\multiput(56.19,32.61)(0.16,-0.48){1}{\line(0,-1){0.48}}
\multiput(56.35,32.13)(0.17,-0.47){1}{\line(0,-1){0.47}}
\multiput(56.51,31.66)(0.18,-0.47){1}{\line(0,-1){0.47}}
\multiput(56.69,31.19)(0.09,-0.23){2}{\line(0,-1){0.23}}
\multiput(56.87,30.72)(0.1,-0.23){2}{\line(0,-1){0.23}}
\multiput(57.07,30.26)(0.1,-0.23){2}{\line(0,-1){0.23}}
\multiput(57.27,29.8)(0.11,-0.23){2}{\line(0,-1){0.23}}
\multiput(57.49,29.34)(0.11,-0.23){2}{\line(0,-1){0.23}}
\multiput(57.71,28.89)(0.12,-0.22){2}{\line(0,-1){0.22}}
\multiput(57.94,28.44)(0.12,-0.22){2}{\line(0,-1){0.22}}
\multiput(58.18,28)(0.12,-0.22){2}{\line(0,-1){0.22}}
\multiput(58.43,27.57)(0.13,-0.22){2}{\line(0,-1){0.22}}
\multiput(58.68,27.13)(0.13,-0.21){2}{\line(0,-1){0.21}}
\multiput(58.95,26.71)(0.14,-0.21){2}{\line(0,-1){0.21}}
\multiput(59.22,26.29)(0.14,-0.21){2}{\line(0,-1){0.21}}
\multiput(59.51,25.87)(0.15,-0.21){2}{\line(0,-1){0.21}}
\multiput(59.8,25.46)(0.15,-0.2){2}{\line(0,-1){0.2}}
\multiput(60.1,25.06)(0.1,-0.13){3}{\line(0,-1){0.13}}
\multiput(60.4,24.66)(0.1,-0.13){3}{\line(0,-1){0.13}}
\multiput(60.72,24.27)(0.11,-0.13){3}{\line(0,-1){0.13}}
\multiput(61.04,23.88)(0.11,-0.13){3}{\line(0,-1){0.13}}
\multiput(61.37,23.5)(0.11,-0.12){3}{\line(0,-1){0.12}}
\multiput(61.71,23.13)(0.11,-0.12){3}{\line(0,-1){0.12}}
\multiput(62.05,22.76)(0.12,-0.12){3}{\line(0,-1){0.12}}
\multiput(62.4,22.4)(0.12,-0.12){3}{\line(1,0){0.12}}
\multiput(62.76,22.05)(0.12,-0.11){3}{\line(1,0){0.12}}
\multiput(63.13,21.71)(0.12,-0.11){3}{\line(1,0){0.12}}
\multiput(63.5,21.37)(0.13,-0.11){3}{\line(1,0){0.13}}
\multiput(63.88,21.04)(0.13,-0.11){3}{\line(1,0){0.13}}
\multiput(64.27,20.72)(0.13,-0.1){3}{\line(1,0){0.13}}
\multiput(64.66,20.4)(0.13,-0.1){3}{\line(1,0){0.13}}
\multiput(65.06,20.1)(0.2,-0.15){2}{\line(1,0){0.2}}
\multiput(65.46,19.8)(0.21,-0.15){2}{\line(1,0){0.21}}
\multiput(65.87,19.51)(0.21,-0.14){2}{\line(1,0){0.21}}
\multiput(66.29,19.22)(0.21,-0.14){2}{\line(1,0){0.21}}
\multiput(66.71,18.95)(0.21,-0.13){2}{\line(1,0){0.21}}
\multiput(67.13,18.68)(0.22,-0.13){2}{\line(1,0){0.22}}
\multiput(67.57,18.43)(0.22,-0.12){2}{\line(1,0){0.22}}
\multiput(68,18.18)(0.22,-0.12){2}{\line(1,0){0.22}}
\multiput(68.44,17.94)(0.22,-0.12){2}{\line(1,0){0.22}}
\multiput(68.89,17.71)(0.23,-0.11){2}{\line(1,0){0.23}}
\multiput(69.34,17.49)(0.23,-0.11){2}{\line(1,0){0.23}}
\multiput(69.8,17.27)(0.23,-0.1){2}{\line(1,0){0.23}}
\multiput(70.26,17.07)(0.23,-0.1){2}{\line(1,0){0.23}}
\multiput(70.72,16.87)(0.23,-0.09){2}{\line(1,0){0.23}}
\multiput(71.19,16.69)(0.47,-0.18){1}{\line(1,0){0.47}}
\multiput(71.66,16.51)(0.47,-0.17){1}{\line(1,0){0.47}}
\multiput(72.13,16.35)(0.48,-0.16){1}{\line(1,0){0.48}}
\multiput(72.61,16.19)(0.48,-0.15){1}{\line(1,0){0.48}}
\multiput(73.09,16.04)(0.48,-0.14){1}{\line(1,0){0.48}}
\multiput(73.57,15.9)(0.49,-0.13){1}{\line(1,0){0.49}}
\multiput(74.06,15.78)(0.49,-0.12){1}{\line(1,0){0.49}}
\multiput(74.55,15.66)(0.49,-0.11){1}{\line(1,0){0.49}}
\multiput(75.04,15.55)(0.49,-0.1){1}{\line(1,0){0.49}}
\multiput(75.53,15.45)(0.49,-0.09){1}{\line(1,0){0.49}}
\multiput(76.03,15.36)(0.5,-0.08){1}{\line(1,0){0.5}}
\multiput(76.52,15.28)(0.5,-0.07){1}{\line(1,0){0.5}}
\multiput(77.02,15.21)(0.5,-0.06){1}{\line(1,0){0.5}}
\multiput(77.52,15.15)(0.5,-0.05){1}{\line(1,0){0.5}}
\multiput(78.02,15.1)(0.5,-0.04){1}{\line(1,0){0.5}}
\multiput(78.52,15.06)(0.5,-0.03){1}{\line(1,0){0.5}}
\multiput(79.02,15.03)(0.5,-0.02){1}{\line(1,0){0.5}}
\multiput(79.52,15.01)(0.5,-0.01){1}{\line(1,0){0.5}}
\put(80.03,15){\line(1,0){0.5}}
\multiput(80.53,15)(0.5,0.01){1}{\line(1,0){0.5}}
\multiput(81.03,15.01)(0.5,0.02){1}{\line(1,0){0.5}}
\multiput(81.53,15.03)(0.5,0.03){1}{\line(1,0){0.5}}
\multiput(82.04,15.06)(0.5,0.04){1}{\line(1,0){0.5}}
\multiput(82.54,15.1)(0.5,0.05){1}{\line(1,0){0.5}}
\multiput(83.04,15.15)(0.5,0.06){1}{\line(1,0){0.5}}
\multiput(83.54,15.21)(0.5,0.07){1}{\line(1,0){0.5}}
\multiput(84.03,15.28)(0.5,0.08){1}{\line(1,0){0.5}}
\multiput(84.53,15.36)(0.49,0.09){1}{\line(1,0){0.49}}
\multiput(85.02,15.45)(0.49,0.1){1}{\line(1,0){0.49}}
\multiput(85.52,15.55)(0.49,0.11){1}{\line(1,0){0.49}}
\multiput(86.01,15.66)(0.49,0.12){1}{\line(1,0){0.49}}
\multiput(86.5,15.78)(0.49,0.13){1}{\line(1,0){0.49}}
\multiput(86.98,15.9)(0.48,0.14){1}{\line(1,0){0.48}}
\multiput(87.47,16.04)(0.48,0.15){1}{\line(1,0){0.48}}
\multiput(87.95,16.19)(0.48,0.16){1}{\line(1,0){0.48}}
\multiput(88.42,16.35)(0.47,0.17){1}{\line(1,0){0.47}}
\multiput(88.9,16.51)(0.47,0.18){1}{\line(1,0){0.47}}
\multiput(89.37,16.69)(0.23,0.09){2}{\line(1,0){0.23}}
\multiput(89.84,16.87)(0.23,0.1){2}{\line(1,0){0.23}}
\multiput(90.3,17.07)(0.23,0.1){2}{\line(1,0){0.23}}
\multiput(90.76,17.27)(0.23,0.11){2}{\line(1,0){0.23}}
\multiput(91.21,17.49)(0.23,0.11){2}{\line(1,0){0.23}}
\multiput(91.66,17.71)(0.22,0.12){2}{\line(1,0){0.22}}
\multiput(92.11,17.94)(0.22,0.12){2}{\line(1,0){0.22}}
\multiput(92.55,18.18)(0.22,0.12){2}{\line(1,0){0.22}}
\multiput(92.99,18.43)(0.22,0.13){2}{\line(1,0){0.22}}
\multiput(93.42,18.68)(0.21,0.13){2}{\line(1,0){0.21}}
\multiput(93.85,18.95)(0.21,0.14){2}{\line(1,0){0.21}}
\multiput(94.27,19.22)(0.21,0.14){2}{\line(1,0){0.21}}
\multiput(94.69,19.51)(0.21,0.15){2}{\line(1,0){0.21}}
\multiput(95.1,19.8)(0.2,0.15){2}{\line(1,0){0.2}}
\multiput(95.5,20.1)(0.13,0.1){3}{\line(1,0){0.13}}
\multiput(95.9,20.4)(0.13,0.1){3}{\line(1,0){0.13}}
\multiput(96.29,20.72)(0.13,0.11){3}{\line(1,0){0.13}}
\multiput(96.68,21.04)(0.13,0.11){3}{\line(1,0){0.13}}
\multiput(97.06,21.37)(0.12,0.11){3}{\line(1,0){0.12}}
\multiput(97.43,21.71)(0.12,0.11){3}{\line(1,0){0.12}}
\multiput(97.79,22.05)(0.12,0.12){3}{\line(1,0){0.12}}
\multiput(98.15,22.4)(0.12,0.12){3}{\line(0,1){0.12}}
\multiput(98.5,22.76)(0.11,0.12){3}{\line(0,1){0.12}}
\multiput(98.85,23.13)(0.11,0.12){3}{\line(0,1){0.12}}
\multiput(99.19,23.5)(0.11,0.13){3}{\line(0,1){0.13}}
\multiput(99.52,23.88)(0.11,0.13){3}{\line(0,1){0.13}}
\multiput(99.84,24.27)(0.1,0.13){3}{\line(0,1){0.13}}
\multiput(100.15,24.66)(0.1,0.13){3}{\line(0,1){0.13}}
\multiput(100.46,25.06)(0.15,0.2){2}{\line(0,1){0.2}}
\multiput(100.76,25.46)(0.15,0.21){2}{\line(0,1){0.21}}
\multiput(101.05,25.87)(0.14,0.21){2}{\line(0,1){0.21}}
\multiput(101.33,26.29)(0.14,0.21){2}{\line(0,1){0.21}}
\multiput(101.61,26.71)(0.13,0.21){2}{\line(0,1){0.21}}
\multiput(101.87,27.13)(0.13,0.22){2}{\line(0,1){0.22}}
\multiput(102.13,27.57)(0.12,0.22){2}{\line(0,1){0.22}}
\multiput(102.38,28)(0.12,0.22){2}{\line(0,1){0.22}}
\multiput(102.62,28.44)(0.12,0.22){2}{\line(0,1){0.22}}
\multiput(102.85,28.89)(0.11,0.23){2}{\line(0,1){0.23}}
\multiput(103.07,29.34)(0.11,0.23){2}{\line(0,1){0.23}}
\multiput(103.28,29.8)(0.1,0.23){2}{\line(0,1){0.23}}
\multiput(103.49,30.26)(0.1,0.23){2}{\line(0,1){0.23}}
\multiput(103.68,30.72)(0.09,0.23){2}{\line(0,1){0.23}}
\multiput(103.87,31.19)(0.18,0.47){1}{\line(0,1){0.47}}
\multiput(104.04,31.66)(0.17,0.47){1}{\line(0,1){0.47}}
\multiput(104.21,32.13)(0.16,0.48){1}{\line(0,1){0.48}}
\multiput(104.37,32.61)(0.15,0.48){1}{\line(0,1){0.48}}
\multiput(104.51,33.09)(0.14,0.48){1}{\line(0,1){0.48}}
\multiput(104.65,33.57)(0.13,0.49){1}{\line(0,1){0.49}}
\multiput(104.78,34.06)(0.12,0.49){1}{\line(0,1){0.49}}
\multiput(104.9,34.55)(0.11,0.49){1}{\line(0,1){0.49}}
\multiput(105.01,35.04)(0.1,0.49){1}{\line(0,1){0.49}}
\multiput(105.11,35.53)(0.09,0.49){1}{\line(0,1){0.49}}
\multiput(105.2,36.03)(0.08,0.5){1}{\line(0,1){0.5}}
\multiput(105.28,36.52)(0.07,0.5){1}{\line(0,1){0.5}}
\multiput(105.35,37.02)(0.06,0.5){1}{\line(0,1){0.5}}
\multiput(105.41,37.52)(0.05,0.5){1}{\line(0,1){0.5}}
\multiput(105.46,38.02)(0.04,0.5){1}{\line(0,1){0.5}}
\multiput(105.5,38.52)(0.03,0.5){1}{\line(0,1){0.5}}
\multiput(105.53,39.02)(0.02,0.5){1}{\line(0,1){0.5}}
\multiput(105.55,39.52)(0.01,0.5){1}{\line(0,1){0.5}}

\end{picture}

}

\newcommand\f{\frac}

\newcommand\bn{{\bf n}}

\begin{document}

\baselineskip 24pt

\begin{center}

{\Large \bf All Order Classical Electromagnetic Soft Theorems}

\end{center}

\vskip .6cm
\medskip

\vspace*{4.0ex}

\baselineskip=18pt

\centerline{\large \rm Debanjan Karan$^1$, Babli Khatun$^1$, Biswajit Sahoo$^2$ 
and
Ashoke Sen$^1$}

\vspace*{4.0ex}

\centerline{\large \it $^1$International Centre for Theoretical Sciences - TIFR 
}
\centerline{\large \it  Bengaluru - 560089, India}

\centerline{\large \it $^2$Department of Mathematics, King's College London}
 \centerline{\large \it Strand, London WC2R 2LS, United Kingdom}

\vspace*{1.0ex}
\centerline{\small E-mail:  debanjan.karan,babli.khatun,ashoke.sen@icts.res.in, 
biswajit.sahoo@kcl.ac.uk}

\vspace*{5.0ex}

\centerline{\bf Abstract} \bigskip

If a set of charged objects collide in space and the fragments disperse, then this process
will emit electromagnetic waves. Classical soft photon theorem determines the constant term and the
leading power 
law fall-off of the wave-form at
late and early times
in terms of only the momenta and charges of the incoming and outgoing objects. 
In this paper we 
determine an infinite set of subleading terms in the late and early time
expansion of the wave-form, which also depend only on the momenta and charges
of the 
incoming and outgoing particles. For two-particle scattering, we derive a resummed low-frequency electromagnetic wave-form, as well as the resummed wave-form at early and late times. 
In this analysis we ignore the effect of long range gravitational interaction, 
but our result is unaffected by any other short range interactions
among the objects.

\vfill \eject

\tableofcontents

\sectiono{Introduction and summary} \label{s0}

Consider a scattering event 
where a set of charged particles come together, interact via electromagnetic and other
short range forces and disperse, possibly exchanging mass, charge and other quantum
numbers, fusing or splitting into fragments. During this process there will be emission
of classical electromagnetic waves. The wave-form 
depends on the detailed acceleration of the 
particles during the scattering and hence on the various long and
short range forces that operated during this period. However at late and early times,
when the particles are almost free, the situation simplifies since the only forces that
act on the particles are long range electromagnetic and gravitational
forces. If we ignore the effect of gravitational forces,  the acceleration of the
particles  and the associated electromagnetic radiation will depend on the various electromagnetic properties of  the particles
{\it e.g.} charges and various multipole moments, as well as the parameters that
characterize the asymptotic trajectories, {\it e.g.} the momenta and relative 
positions of the trajectories.

In \cite{1912.06413} and \cite{2008.04376} (see also \cite{1804.09193,1806.01872,1808.03288,AtulBhatkar:2020hqz})
it was shown that if $u$ is the retarded time measured by a detector placed
at $\II^+$, with $u=0$ representing the time when the peak of the signal reaches
the detector, then the coefficients of 
the  constant, $1/u$ and $u^{-2}\ln|u|$ 
terms in the wave-form for large positive $u$ and large
negative $u$ are determined solely in terms of the momenta and charges of the incoming
and the outgoing particles.
In this paper we show that the same is true 
for the coefficients
of the $u^{-n-1} (\ln |u|)^n$ terms for all $n\ge 0$ and derive the explicit expressions
for the coefficients of these terms. In this analysis we ignore the presence of the long
range gravitational interaction between the particles, but our results are insensitive to
any other short range forces that act between the particles during scattering, including
possible fusion and splitting of the particles. It is expected that the inclusion of
gravitational interaction will not spoil the universality of these coefficients but will
change their actual expression (see {\it e.g.} \cite{1912.06413}).

There are several motivations behind this analysis. First of all, similar universal results
exist for the coefficients of the constant, 
$u^{-1}$ and $u^{-2} \ln |u|$ terms in the
gravitational wave-form produced by the scattering of particles.
This 
has been verified by explicit computation in many
examples\cite{1901.10986,2007.02077,2211.13120, DiVecchia:2023frv,
2309.14925,2312.07452,2402.06361,2402.06533,2402.06604,
2407.02076,2407.04128,2408.07329}.
In a recent 
paper \cite{2407.04128} by Alessio, Di Vecchia and Heissenberg, 
a generalization of these formulae\ for the $u^{-n-1} (\ln|u|)^n$ terms for all $n$ was
conjectured and a proof of this conjecture has been provided in \cite{ 2408.07329} under the probe approximation.\footnote{Earlier in \cite{2008.04376}, 
 motivated by the infinite order tree-level soft theorems \cite{1801.05528,1802.03148},
 a similar conjecture for the electromagnetic and gravitational wave-forms at order $u^{-n-1} (\ln|u|)^n$  was proposed up to an undetermined gauge-invariant contribution.} It is in principle possible to directly test this conjecture by generalizing
the analysis described in this paper to the gravitational wave-form, but the
analysis of the electromagnetic wave-form offers a simpler setting. Our result shows that
for scattering that involves at most two initial particles and at most two final particles,
the result for the electromagnetic wave-form follows a pattern analogous to the one 
conjectured in \cite{2407.04128} for the gravitational wave-form, but for
three or more initial and / or final particles the result takes a much more complicated form.
This suggests that even for the gravitational wave-form the conjecture of 
\cite{2407.04128} is likely to be true for the scatterings involving two or
less particles in the initial and final states, but is likely to be modified for scattering 
involving three or more initial and / or final particles.

A second motivation comes from the interrelation between the quantum version of soft theorems relating S-matrices, and the classical version discussed here. For example, the late-time electromagnetic wave-form at orders $u^{-1}$ and $u^{-2} \ln |u|$ is related to the quantum soft photon theorem at orders $\ln\omega$ and $\omega(\ln\omega)^2$ respectively for soft photon energy $\omega$, as derived in \cite{1808.03288,2008.04376,2308.16807} by analyzing one- and two-loop QED S-matrices. Hence, it is anticipated that an analogous study of $(n+1)$-loop QED S-matrices should enable one to derive the soft photon theorem at order $\omega^{n}(\ln\omega)^{n+1}$ which, in the proper classical limit, should be related to the electromagnetic wave-form at order $u^{-n-1} (\ln u)^{n}$ derived in this paper.

A third motivation comes from the relation between the classical
electromagnetic wave-form and asymptotic symmetries. In particular it was shown
recently \cite{2412.16149,2412.16142}  (see \cite{1903.09133,1912.10229,2205.11477,2309.11220,2403.13053,2407.07978}
for earlier work on this subject)
that the coefficient of the $1/u$ term in the electromagnetic and gravitational
wave-forms at late and early times can be explained in terms of asymptotic symmetries
and their associated conserved charges. This naturally leads to the question as to
whether the coefficients of the $u^{-n-1} (\ln u)^{n}$ terms can also be explained
using some asymptotic symmetries and the associated conserved charges.

We now summarize the main results of this paper. 
We take the origin of space-time to be somewhere inside the region where the
scattering takes place. This only defines the location of the origin up to shifts
of order unity, but this will be sufficient to state our result. Let $\vec x$ be the spatial
coordinate of the detector in this coordinate system. We define 
\be
R\equiv |\vec x|, \qquad \hat n \equiv {\vec x\over R},  \qquad
\bn \equiv (1, \hat n), \qquad u \equiv 
x^0 - R\, .
\ee
$u$ is the retarded time at the detector.
Then the peak of the signal reaches the detector at $u\sim 1$ and our focus will be
on the order $R^{-1}$ term in the
electromagnetic field $A_\mu$ at the detector at large $|u|$. 
Let us first consider the wave-form
at large positive $u$, which is controlled by the outgoing
particle trajectories at late time.
The relevant part of the trajectory of the $a$-th outgoing 
particle can be labelled as
\be \label{eansatz}
X_{a}^\mu(\tau_a)= \f{p_a^\mu}{m_a} \, \tau_a 
+ \left( C_1^{(a)\mu} \ln\tau_a +\OO(1)\right)
-\sum_{r=1}^\infty  \left( C_{r+1}^{(a)\mu}\, \f{(\ln\tau_a)^{r}}{\tau_a^{r}}+\OO(\tau_a^{-r}(\ln\tau_a)^{r-1})\right) +\cdots ,\label{eq:trajectory_patternb}
\ee
where $\tau_a$ is the proper time of the $a$'th particle, $p_a^\mu$ is its momentum,
$m_a$ is its mass, $q_a$ is its charge and
$C_r^{(a)\mu}$'s are constants.
Note that the terms that we ignore are not necessarily smaller that 
the terms that we keep, 
{\it e.g.}  for large $\tau_a$,
the $\tau_a^{-r+1}(\ln\tau_a)^{r-2}$ term that we ignore is larger that the 
$\tau_a^{-r}(\ln\tau_a)^r$ term that we keep.
Nevertheless these are the terms that will be relevant for determining the
required terms in the wave-form. 
We find that the equations of motion determine the coefficients $C_r^{(a)\mu}$'s
to be,
\be\label{eC1expint}
C_1^{(a)\mu}=\sum_{b\neq a} \f{q_a q_b}{4\pi} \f{p_b^2}{\left[ (p_a.p_b)^2-p_a^2 p_b^2\right]^{\f{3}{2}}}\left(p_a^2 p_b^\mu -p_a.p_b p_a^\mu\right),
\ee
\ben \label{eCrexpint}
&&C_{n+1}^{(a)\mu}\non\\
&&=\f{m_a^n}{2^nn(n+1)}\sum_{b\neq a}\f{q_a q_b}{4\pi} \f{p_b^2}{\left[(p_a.p_b)^2-p_a^2 p_b^2\right]^\f{3}{2}} 
\Bigg[ \left(p_a^2 p_b^\mu -p_a.p_b p_a^\mu\right)\non\\
&&\times \sum_{r=\f{n}{2}\ 
{\rm for}\, n \, {\rm even}\atop
 r=\f{n+1}{2}\ {\rm for} \, n\, {\rm odd}}^n \f{(-1)^r\ (2r+1)!}{ r! (2r-n)! (n-r)!}  \f{1}{\left((p_a.p_b)^2-p_a^2 p_b^2\right)^r}\non\\
 &&\times \left(  p_b^2\, p_a.C_1^{(b)}
+p_a.p_b p_b.C_1^{(a)}\right)^{2r-n}   \left(  \left(p_b.C_1^{(a)}\right)^2-p_b^2 \left(C_1^{(b)}-C_1^{(a)}\right)^2 \right)^{n-r}\non\\
&& +\ 2\left( p_a.p_b\left(C_1^{(b)\mu}-C_1^{(a)\mu}\right) 
- p_a.\, C_1^{(b)}\,  p_b^\mu  \right) \non\\ &&
\times \ \sum_{r=\f{n}{2}\ {\rm for} \, n \, {\rm even}\atop r=\f{n-1}{2}\ {\rm for}\,
n \, {\rm odd}}^{n-1} \f{(-1)^r\ (2r+1)!}{ r! (2r+1-n)! (n-r-1)!} \non\\
&&\times \f{1}{\left((p_a.p_b)^2-p_a^2 p_b^2\right)^r}
\left(   p_b^2\, p_a.C_1^{(b)} 
+p_a.p_b p_b.C_1^{(a)} \right)^{2r+1-n}\non\\
&&\times  \left(  \left(p_b.C_1^{(a)}\right)^2-p_b^2 \left(C_1^{(b)}-C_1^{(a)}\right)^2\right)^{n-r-1}\Bigg]\, ,  \qquad \hbox{for $n\ge 1$}\, .
\een
This ignores the effect of any other long range force among the particles (e.g. gravitational or dilaton mediated forces), but
any short range force that falls off faster than the inverse square law will not affect
the result.

In terms of the above trajectory coefficients, the electromagnetic field $A_\mu$ at the detector at late retarded time $u$ is given by 
\ben\label{eq:log_n_waveform}
&&A^{\mu}(u,R\hat{n})\non\\
&\simeq  &- \f{1}{4\pi R}\sum_{a} \f{q_ap_a^\mu}{\mathbf{n}.p_a} +
 \f{1}{4\pi R}\sum_{a} \f{q'_ap_a^{\prime\mu}}{\mathbf{n}.p'_a}\non\\
&&+  \f{1}{4\pi R}\sum_{n=1}^\infty (-1)^n \f{(\ln u)^{n-1}}{u^n}\sum_{a}\f{q_a}{p_a.\mathbf{n}}\Bigg[ \left( \mathbf{n}.C_1^{(a)}\right)^{n-1} \left( \mathbf{n}.C_1^{(a)} p_a^\mu -\mathbf{n}.p_a  C_1^{(a)\mu} \right) \non\\
&&+\sum_{r=1}^{n-1} \f{(n-1)!}{(r-1)! (n-r-1)!} \left(\mathbf{n}.C_{r+1}^{(a)} p_a^\mu-\mathbf{n}.p_a C_{r+1}^{(a)\mu} \right) \left(\f{\mathbf{n}.p_a}{m_a}\right)^r  \left( \mathbf{n}.C_1^{(a)}\right)^{n-r-1}\Bigg],
\een
where $\{p'_a\}$, $\{q'_a\}$ and $\{m'_a\}$ denote the momenta, charges and masses
of the incoming particles.
Note that the first line of the electromagnetic wave-form above is $u$ independent and is known as electromagnetic memory. This is expressed in an observer's frame where the electromagnetic wave-form vanishes at infinitely early times. In the above expression, under the $\simeq$ sign, we have retained the time-dependent part of the wave-form, which behaves like $u^{-n}(\ln u)^{n-1}$ for $n \geq 1$. This represents the leading-log late-time contribution at order $u^{-n}$ in the large $u$ expansion and ignores the contributions at orders $u^{-n}(\ln u)^m$ for $m \leq n-2$.

The result for large negative $u$ can be obtained from \refb{eq:log_n_waveform} after substituting \eqref{eC1expint} and \eqref{eCrexpint} as follows. First
of all the sum over $a$ will now denote the sum over incoming particles and 
$p_a^\mu$, $q_a$ and $m_a$ will have to be replaced by the
momentum $p_a^{\prime\mu}$, charge $q'_a$ and mass $m'_a$ 
of the incoming particles. Also all factors of $\ln u$ and $u$ in 
\refb{eq:log_n_waveform} will now be
replaced by $\ln |u|$ and $|u|$ respectively. Additionally, in our observer's frame, we need to remove the constant term in the first line on the right-hand side.
 The late and early time wave-forms following 
 from \eqref{eq:log_n_waveform} at order $u^{-1}$ and 
 $u^{-2}\ln u$ agree with the results in \cite{1912.06413} and \cite{2008.04376} respectively. 

It is also instructive to present the result in terms of the Fourier transform
\be
\wt A^\mu (\omega, \vec x) \equiv \int_{-\infty}^\infty du\, e^{i\omega u} A_\mu(t, \vec x)\, .
\ee
If we express the large $|u|$ behaviour of $A_\mu$ as
\ben
A_\mu &\simeq& A_{(0)\mu} - \sum_{r=1}^\infty \,  A_{(r)\mu} \, u^{-r} (\ln u)^{r-1} \qquad
\hbox{for large $u$} \, , \nonumber \\
&\simeq& A'_{(0)\mu} - \sum_{r=1}^\infty \,  A'_{(r)\mu} \, u^{-r} (\ln |u|)^{r-1} \qquad
\hbox{for large $-u$}\, ,
\een
then this
translates to the following small $\omega$
behaviour:\footnote{ The $i\eps$ prescription is irrelevant in the 
$(\omega\pm i\eps)^{r-1}$ factor except for $r=0$.}
\be\label{eFT}
\wt A_\mu (\omega, \vec x) \simeq \sum_{r=0}^\infty \,  (-i)^{r-1}\, {1\over r!}\,
 \left[A_{(r)\mu} \, (\omega+i\eps)^{r-1} \, \{\ln(\omega+i\eps)\}^r 
- A'_{(r)\mu} \, (\omega-i\eps)^{r-1} \,\{\ln(\omega-i\eps)\}^r\right] \, .
\ee
Note that the terms that we have kept in the expression for the wave-form
are often smaller than the terms we have not kept, {\it e.g.}
in \refb{eFT} the $\omega^{r-1} (\ln\omega)^{r}$ term is small compared to the
$\omega^{r-2} (\ln\omega)^{r-2}$ term for small $\omega$. 
However, as discussed in \cite{2408.08851}, by
applying appropriate differential operator on the wave-form we can ensure that the
dominant contribution comes from the term proportional to 
$\omega^{r-1} (\ln\omega)^{r}$ in $\wt A_\mu$. This terms will be referred to leading-log contributions to $\wt A_\mu$ in small $\omega$ expansion, as at order $\omega^{r-1}$ this is the leading singular term. Conversely we can regard this as the leading term for small $\omega$ that carries $r$ powers of $\ln\omega$.

When the number of final state particles is two, then 
\refb{eCrexpint} simplifies to:
\be \label{e1.11}
C_{r+1}^{(a)\mu}= \f{m_a^r}{r}\,     
\left( \f{q_a q_b}{4\pi} \f{ p_b^2 }{\left[(p_a.p_b)^2-p_a^2 p_b^2\right]^\f{3}{2}}
\right)^{r+1}\ \left[p_a.(p_a+p_b)\right]^r\,  \left(p_a^2 p_b^\mu -p_a.p_b p_a^\mu\right),
\qquad b\ne a.
\ee
Substituting \eqref{eC1expint} and \eqref{e1.11} in the trajectory ansatz \eqref{eq:trajectory_patternb}, we find the following resummed trajectory 
\ben\label{eq:trajectory_resummed}
X_a^{\mu}(\tau_a)&=&\f{p_a^\mu}{m_a}\tau_a + \left( \f{q_a q_b}{4\pi} \f{ p_b^2 }{\left[(p_a.p_b)^2-p_a^2 p_b^2\right]^\f{3}{2}}
\right) \left(p_a^2 p_b^\mu -p_a.p_b p_a^\mu\right)\non\\
&&\times \ln\left(\tau_a -m_a p_a.(p_a+p_b)   \f{q_a q_b}{4\pi}\f{ p_b^2 }{\left[(p_a.p_b)^2-p_a^2 p_b^2\right]^\f{3}{2}}  \ln\tau_a+\OO(\tau_a^0) \right)\non\\
&&+\cdots, \qquad b\ne a.
\een

Consequently the expression for the late retarded time wave-form given in
\refb{eq:log_n_waveform} also simplifies to:
\ben\label{eq:log_n_waveform_2particleint}
&&A^{\mu}(u,R\hat{n})\non\\
&\simeq &- \f{1}{4\pi R}\sum_{a=1,2} \f{q_ap_a^\mu}{\mathbf{n}.p_a} +
 \f{1}{4\pi R}\sum_{a=1,2} \f{q'_ap_a^{\prime\mu}}{\mathbf{n}.p'_a}\non\\
 &&+\f{1}{4\pi R}\sum_{n=1}^\infty (-1)^{n} \f{(\ln u)^{n-1}}{u^n}
\sum_{a,b \neq a}
\left(\sigma_{ab}^{out}\right)^{n}\f{q_a}{p_a.\mathbf{n}}\left(\mathbf{n}.p_b p_a^\mu -\mathbf{n}.p_a p_b^\mu\right) \left( (p_a+p_b).\mathbf{n}\right)^{n-1},
\een
where
\be\label{edefsigmaabout}
\sigma_{ab}^{out}\equiv \f{q_a q_b}{4\pi}\f{p_a^2 p_b^2}{\left[(p_a.p_b)^2-p_a^2 p_b^2\right]^\f{3}{2}}, \qquad
\sigma_{ab}^{in}\equiv \f{q'_a q'_b}{4\pi}\f{p_a^{\prime 2} p_b^{\prime 2}}{\left[(p'_a.p'_b)^2
-p_a^{\prime 2} p_b^{\prime 2}\right]^\f{3}{2}}\, .
\ee
After performing the sum over $n$, the electromagnetic wave-form at late retarded time provided in \eqref{eq:log_n_waveform_2particleint} becomes
\ben\label{eq:resummed_u_2particle_late}
A^{\mu}(u,R\hat{n})
&\simeq &- \f{1}{4\pi R}\sum_{a=1,2} \f{q_ap_a^\mu}{\mathbf{n}.p_a} +
 \f{1}{4\pi R}\sum_{a=1,2} \f{q'_ap_a^{\prime\mu}}{\mathbf{n}.p'_a}\non\\
 &&-\f{1}{4\pi R}\sigma_{12}^{out} \, \f{1}{u} \left(1+\sigma_{12}^{out}\, (p_1+p_2)\cdot\mathbf{n}\, \f{\ln u}{u}\right)^{-1}D_{out}^\mu,
\een
where
\ben
D^\mu_{out}&\equiv & \left(\frac{q_1p_1^\mu}{p_1\cdot\mathbf{n}}p_2\cdot\mathbf{n}+\frac{q_2p_2^\mu}{p_2\cdot\mathbf{n}}p_1\cdot\mathbf{n}-q_1p_2^\mu -q_2p_1^\mu\right).\label{eq:D_out}
\een
Similar simplification occurs for early retarded time wave-form when the number of
incoming particles is two. The resummed electromagnetic wave-form at early retarded time in the observer's frame reads
\ben\label{eq:resummed_u_2particle_early}
&&A^{\mu}(u,R\hat{n})
\simeq \f{1}{4\pi R}\sigma_{12}^{in} \, \f{1}{u} \left(1-\sigma_{12}^{in}\, (p'_1+p'_2)\cdot\mathbf{n}\, \f{\ln |u|}{u}\right)^{-1}D_{in}^\mu,
\een
where
\ben
D^\mu_{in}&\equiv & \left(\frac{q^\prime_1p_1^{\prime\mu}}{p^\prime_1\cdot\mathbf{n}}p^\prime_2\cdot\mathbf{n}+\frac{q^\prime_2p_2^{\prime\mu}}{p^\prime_2\cdot\mathbf{n}}p^\prime_1\cdot\mathbf{n}-q^\prime_1p_2^{\prime\mu} -q^\prime_2p_1^{\prime\mu}\right).\label{eq:D_in}
\een

When both, the number of incoming particles and the number of outgoing particles
are two, then in the observer's frame 
the Fourier transformed wave-form given in \refb{eFT} also
takes a compact form:
\ben\label{eq:log_n_waveform_2particle_frequency}
&& \wt A^\mu (\omega, \vec x)\non\\
&\simeq&-\f{i}{4\pi R} \f{1}{\omega  +i\eps}\sum_{a=1,2}\left(\f{q_ap_a^\mu}{p_a.\mathbf{n}}-\f{q'_ap_a^{\prime\mu}}{p'_a.\mathbf{n}}\right)\non\\
&&- \f{i}{4\pi R}\sum_{n=1}^\infty \f{1}{n!}
\omega^{n-1}\lbrace \ln(\omega+i\epsilon)\rbrace^n \sum_{a,b\atop b\ne a}  
\left(i\sigma_{ab}^{out}\right)^{n}\f{q_a}{p_a.\mathbf{n}}\left(\mathbf{n}.p_b p_a^\mu -\mathbf{n}.p_a p_b^\mu\right) \left( (p_a+p_b).\mathbf{n}\right)^{n-1} \non\\ &&+
 \f{i}{4\pi R}\sum_{n=1}^\infty \f{1}{n!}
\omega^{n-1}\lbrace \ln(\omega-i\epsilon)\rbrace^n \sum_{a,b\atop b\ne a}  
\left(-i\sigma_{ab}^{in}\right)^{n}\f{q'_a}{p'_a.\mathbf{n}}\left(\mathbf{n}.p'_b 
p_a^{\prime \mu} 
-\mathbf{n}.p'_a p_b^{\prime\mu}\right) \left( (p'_a+p'_b).\mathbf{n}\right)^{n-1} \, .\non\\
\een
After performing the sum over $n$, the resummed electromagnetic wave-form for two-particle scattering becomes
\ben
&& \wt A^\mu (\omega, \vec x)\non\\
&\simeq& - \f{i}{4\pi R} \f{1}{\omega  +i\eps 
}\Bigg[\left(\f{q_1p_1^\mu}{p_1.\mathbf{n}}+\f{q_2p_2^\mu}{p_2.\mathbf{n}}-\f{q'_1p_1^{\prime\mu}}{p'_1.\mathbf{n}}-\f{q'_2p_2^{\prime\mu}}{p'_2.\mathbf{n}}\right)\non\\
&& +\frac{(\omega+i\epsilon)^{i\omega \sigma_{12}^{out}(p_1+p_2)\cdot\mathbf{n}}-1}{(p_1+p_2)\cdot\mathbf{n}}D^\mu_{out}-\frac{(\omega-i\epsilon)^{-i\omega \sigma_{12}^{in}(p^\prime_1+p^\prime_2)\cdot\mathbf{n}}-1}{(p^\prime_1+p^\prime_2)\cdot\mathbf{n}}D^\mu_{in}\Bigg].\label{eq:resummed_em_waveform}
\een
Using the charge and momentum conservation relations, the above resummed expression can be further simplified as follows. Substituting $q'_2 = q_1 + q_2 - q'_1$ and $p'_2 = p_1 + p_2 - p'_1$, the terms inside the parentheses on the first line of the RHS cancel out with the two terms multiplying $(-1)$ in the numerators of the second line of the RHS after substituting \eqref{eq:D_out} and \eqref{eq:D_in}. Finally, the simplified expression becomes:
\ben
&& \wt A^\mu (\omega, \vec x)\non\\
&\simeq& - \f{i}{4\pi R} \f{1}{\omega  +i\eps 
}\Bigg[\frac{(\omega+i\epsilon)^{i\omega \sigma_{12}^{out}(p_1+p_2)\cdot\mathbf{n}}}{(p_1+p_2)\cdot\mathbf{n}}D^\mu_{out}-\frac{(\omega-i\epsilon)^{-i\omega \sigma_{12}^{in}(p^\prime_1+p^\prime_2)\cdot\mathbf{n}}}{(p^\prime_1+p^\prime_2)\cdot\mathbf{n}}D^\mu_{in}\Bigg].\label{eq:resummed_em_waveform_simplified}
\een

\sectiono{Particle trajectories} \label{strajectory}

In this section we shall determine the outgoing particle trajectories at late time.
Since the electromagnetic field at late time is determined by the
form of the outgoing particle trajectories at late time, and the latter in turn is determined
by the electromagnetic field produced by the other particles,
we need to iteratively solve the 
equations of motion for the outgoing
particle trajectories and the electromagnetic field at late
time. In the harmonic gauge these take the form:\footnote{We shall argue below \refb{e2.14} that 
the self-force 
effects are not important to the order that we
shall study the system.}
\ben
m_a\f{d^2 X_a^\mu(\tau_a)}{d\tau_a^2}&=& q_a \f{d X_{a\nu}(\tau_a)}{d\tau_a}
\sum_{b\ne a} F_{(b)}^{\mu\nu}(X_a(\tau_a)), \label{eq:trajectory_newton}\\
\p^\rho \p_\rho A_{(b)\mu}(x)&=&-J_{(b)\mu}(x),\label{eq:Maxwell}
\een
where $A_{(b)\mu}(x)$ and $F_{(b)}^{\mu\nu}(x)$ denote the 
electromagnetic field and field strength produced at $x$ by the
particle $b$ and the current density associated with particle $b$ is given by
\ben
J_{(b)}^\mu(x)=q_b\int_0^\infty d\tau_b \ \delta^{(4)}\left(x-X_b(\tau_b)\right)\f{dX_b^\mu(\tau_b)}{d\tau_b}\, .
\een
The indices $a,b,c$ run over all the outgoing particles.
The solution for $A_{(b)\mu}(x)$ from \eqref{eq:Maxwell} using retarded propagator 
becomes
\ben 
A_{(b)}^{\mu}(x)={q_b\over 2\pi}  \int_0^\infty d\tau_b\,  
\delta_+\left( - \left(x-X_b(\tau_b)\right)^2\right) \f{d X_{b}^{\mu}(\tau_b)}{d\tau_b}\, ,\label{eq:Ab_mu_int_rep}
\een
where $\delta_+$ denotes the usual Dirac delta function multiplying a 
Heaviside theta function $H\left(x^0-X_b^0(\tau_b)\right)$, 
such that we have to choose the zero of the argument for which $x^0>X_b^0(\tau_b)$. 

Now using \eqref{eq:Ab_mu_int_rep}, the field strength $F_{(b)}^{\mu\nu}$ 
produced by particle $b$ reads
\ben \label{e37}
&& F_{(b)}^{\mu\nu}(x) =\p^\mu A_{(b)}^\nu(x) - \p^\nu A_{(b)}^\mu(x) \non\\
&=& -\f{q_b}{2\pi}\int_0^\infty d\tau_b \ \left[\f{\p}{\p\tau_b}\delta_+\left( - \left(x-X_b(\tau_b)\right)^2\right) \right]\non\\
&&\times \f{\left(x^\mu-X_{b}^{\mu}(\tau_b)\right)\f{d X_{b}^{\nu}(\tau_b)}{d\tau_b}- \left(x^\nu-X_{b}^{\nu}(\tau_b)\right)\f{d X_{b}^{\mu}(\tau_b)}{d\tau_b} }{\left(x_\alpha-X_{b\alpha}(\tau_b)\right)\f{dX_b^\alpha(\tau_b)}{d\tau_b}}\non\\
&=& \f{q_b}{2\pi}\int_0^\infty d\tau_b \ \delta_+\left( - \left(x-X_b(\tau_b)\right)^2\right) 
\Bigg[ \f{\left(x^\mu-X_{b}^{\mu}(\tau_b)\right)\f{d^2 X_{b}^{\nu}(\tau_b)}{d\tau_b^2}- \left(x^\nu-
X_{b}^{\nu}(\tau_b)\right)\f{d^2 X_{b}^{\mu}(\tau_b)}{d\tau_b^2} }{\left(x_\alpha-X_{b\alpha}
(\tau_b)\right)\f{dX_b^\alpha(\tau_b)}{d\tau_b}} \non\\
&&-\f{\left(x^\mu-X_{b}^{\mu}(\tau_b)\right)\f{d X_{b}^{\nu}(\tau_b)}{d\tau_b}- \left(x^\nu-
X_{b}^{\nu}(\tau_b)\right)\f{d X_{b}^{\mu}(\tau_b)}{d\tau_b} }{\left(\left(x_\alpha-X_{b\alpha}
(\tau_b)\right)\f{dX_b^\alpha(\tau_b)}{d\tau_b}\right)^2}  \non\\
&& \hskip .5in \times\left( \left(x_\alpha-X_{b\alpha}(\tau_b)\right)\f{d^2X_b^\alpha(\tau_b)}
{d\tau_b^2}-\f{dX_b^\alpha(\tau_b)}{d\tau_b} \f{dX_{b\alpha}(\tau_b)}{d\tau_b} \right) \Bigg]\, , 
\een
where in the second step we have performed an integration by parts, and in the first step used the identity
\ben
\f{\p}{\p x^\nu} \delta_+\left( - \left(x-X_b(\tau_b)\right)^2\right)&=& -2\left(x_\nu-X_{b\nu}(\tau_b)\right)\delta^\prime_+\left( - \left(x-X_b(\tau_b)\right)^2\right)\non\\
&=& -\f{\left(x_\nu-X_{b\nu}(\tau_b)\right)}{\left(x_\alpha-X_{b\alpha}(\tau_b)\right)\f{dX_b^\alpha(\tau_b)}{d\tau_b}}\times \f{\p}{\p\tau_b}\delta_+\left( - \left(x-X_b(\tau_b)\right)^2\right).\label{eq:delta_derivative_identity}
\een

We have to use \refb{e37} to compute the field strength at the location of the particle $a$
so that we can compute the right hand side of \refb{eq:trajectory_newton}.
For this we  use the ansatz \refb{eansatz} for the particle trajectories
and determine $\tau_b$ 
in terms of $\tau_a$ by solving the constraint $(X_a(\tau_a)-X_b(\tau_b))^2=0$
imposed by the
delta function.
At this stage, we need to decide
which terms in the expression for 
$X^\mu_b$ and $X_a^\mu$ 
we need to keep and which terms we can ignore for this analysis. 
Since our goal is to
keep all terms of order $u^{-r-1} (\ln u)^r$ in the wave-form, we formally introduce a
large parameter $\lambda$ and pretend that $u$ and $\ln u$ scale as $\lambda$.
We shall see later that for calculating the wave-form at the detector at
retarded time $u$, the proper
time $\tau_b$ and $\tau_a$ at which we need the trajectories of the $b$-th 
and the $a$-th particle are of order $u$.
Hence $\tau_b$, $\tau_a$ scale as $\lambda$ and $\ln\tau_b\simeq\ln u$,
$\ln\tau_a\simeq\ln u$ also scale as
$\lambda$. Looking back at \refb{eansatz} we see that the terms proportional to
$\tau_a$ and $\ln\tau_a$ are of order $\lambda$ while the coefficients of $C_r^{(a)\mu}$ for $r\geq 2$ are of
order $\lambda^0$. The terms ignored in \eqref{eansatz} scale as $\lambda^{-p}$ for $p\geq 1$. Hence while solving the equation $\left(X_a(\tau_a)-X_b(\tau_b)\right)^2=0$ to
determine $\tau_b$ for a given $\tau_a$, 
we shall keep only the first two terms in the
expression for $X_b(\tau_b)$ and $X_a(\tau_a)$:
\be\label{etrajapp}
X_c^\mu(\tau_c) = {p_c^\mu\over m_c}\,  \tau_c + C_1^{(c)\mu}\ln \tau_c + \OO(\lambda^{0}), \qquad
\hbox{for $c=a,b$}\, .
\ee
Using this the $(X_b(\tau_b)-X_a(\tau_a))^2=0$ equation reduces to
\ben
&&\tau_a^2 +\tau_b^2+2\f{p_a.p_b}{m_a m_b}\tau_a\tau_b +2\f{p_a.\left(C_1^{(b)}-C_1^{(a)}\right)}{m_a}\tau_a\ln\tau_a -2\f{p_b.\left(C_1^{(b)}-C_1^{(a)}\right)}{m_b}\tau_b\ln\tau_a\non\\
&&-\left( C_1^{(b)}-C_1^{(a)}\right)^2 (\ln\tau_a)^2 +\OO(\lambda)=0.
\een
Note that in the LHS of the above expression, we only kept terms from $-\left(X_a(\tau_a)-X_b(\tau_b)\right)^2$ that scale like order $\lambda^2$ for large value of $\tau_a$. This allowed us to replace $\ln\tau_b$ with $\ln\tau_a$ for the trajectory ansatz of $b$'th particle, since for $\tau_b\sim \tau_a$, $\ln\tau_b= 
\ln\tau_a+\OO(\lambda^0)$. The solution reads
\ben\label{eq:tau_ba_star}
\tau_b &=& \tau^\star_{ba}+\OO(\lambda^0)\, ,\non\\
\tau^\star_{ba}&\equiv& -\f{p_a.p_b}{m_a m_b}\tau_a +\f{p_b.\left(C_1^{(b)}-C_1^{(a)}\right)}
{m_b}\ln\tau_a -\left[-\tau_a^2 - \f{2p_a.\left(C_1^{(b)}-C_1^{(a)}\right)}{m_a}
\tau_a\ln\tau_a\right.\non\\
&&\left. +\left( C_1^{(b)}-C_1^{(a)}\right)^2 (\ln\tau_a)^2+\left(\f{p_a.p_b}{m_a m_b}\tau_a -
\f{p_b.\left(C_1^{(b)}-C_1^{(a)}\right)}{m_b}\ln\tau_a\right)^2\right]^\f{1}{2}\, ,
\een
where we have chosen the branch with lower value of $\tau_b$ due to
the Heaviside theta function $H\left(X_a^0(\tau_a)-X_b^0(\tau_b)\right)$ associated with the retarded Green's function. 
As an internal consistency check on our approximation scheme, we see that as long
as $\tau_a$ is of order $\lambda$, $\tau_b$ is also of order $\lambda$,
in agreement with our
earlier assumption.
Using \refb{eq:tau_ba_star} we get
\be
\delta_+ \left(- (X_a(\tau_a)-X_b(\tau_b))^2\right)
=\f{1}{2\, \Big{|} \left(X_{a}(\tau_a)-X_{b}(\tau_b)\right)\cdot\f{dX_b(\tau_b)}{d\tau_b}\Big{|}}_{\tau_b=\tau_{ba}^\star} \delta(\tau_b - \tau_{ba}^*)\, .
\ee
After substituting this into \eqref{e37}, 
we get
\ben\label{eq:F_b}
F_{(b)}^{\mu\nu}\left(X_a(\tau_a)\right)&=& \f{q_b}{4\pi}\f{1}{\Big{|} \left(X_{a}(\tau_a)-X_{b}(\tau_b)\right)\cdot\f{dX_b(\tau_b)}{d\tau_b}\Big{|}}_{\tau_b=\tau_{ba}^\star} \non\\
&&\times \Bigg[ \f{\left(X_a^\mu(\tau_a)-X_{b}^{\mu}(\tau_b)\right)\f{d^2 X_{b}^{\nu}(\tau_b)}{d\tau_b^2}- \left(X_a^\nu(\tau_a)-X_{b}^{\nu}(\tau_b)\right)\f{d^2 X_{b}^{\mu}(\tau_b)}{d\tau_b^2} }{ \left(X_{a}(\tau_a)-X_{b}(\tau_b)\right)\cdot\f{dX_b(\tau_b)}{d\tau_b}} \non\\
&&-\f{\left(X_a^\mu(\tau_a)-X_{b}^{\mu}(\tau_b)\right)\f{d X_{b}^{\nu}(\tau_b)}{d\tau_b}- \left(X_a^\nu(\tau_a)-X_{b}^{\nu}(\tau_b)\right)\f{d X_{b}^{\mu}(\tau_b)}{d\tau_b} }{\left( \left(X_{a}(\tau_a)-X_{b}(\tau_b)\right)\cdot\f{dX_b(\tau_b)}{d\tau_b}\right)^2} \non\\
&&\times \left( \left(X_{a}(\tau_a)-X_{b}(\tau_b)\right)\cdot\f{d^2X_b(\tau_b)}{d\tau_b^2}-\f{dX_b(\tau_b)}{d\tau_b} \cdot \f{dX_{b}(\tau_b)}{d\tau_b} \right) \Bigg]_{\tau_b=\tau_{ba}^\star}.
\een
Using \refb{etrajapp} this can be simplified to
\ben \label{e2.11}
&&F_{(b)}^{\mu\nu}\left(X_a(\tau_a)\right)\non\\
&= & -\f{q_b}{4\pi}\f{1}{\tau_a^2}  {m_a^3 m_b^3 \over ((p_a.p_b)^2 - m_a^2 m_b^2)^{3/2}}
 \non\\ &&
\left[ 1+2m_a  \f{ p_b^2p_a.\left(C_1^{(b)}-C_1^{(a)}\right) -p_a.p_b p_b.\left(C_1^{(b)}-C_1^{(a)}\right) }{(p_a.p_b)^2 -p_a^2 p_b^2}\, \xi_a \right.\non\\
&&\hskip .5in \left.+m_a^2 \f{  \left(p_b.\left(C_1^{(b)}-C_1^{(a)}\right)\right)^2-p_b^2 \left(C_1^{(b)}-C_1^{(a)}\right)^2 }{(p_a.p_b)^2 -p_a^2 p_b^2}\, \xi_a^2\right]^{-\f{3}{2}}  \non\\ &&
  \Bigg[{\f{p_a^\mu p_b^\nu -p_a^\nu p_b^\mu}{m_a m_b}  + \left(C_1^{(a)\mu}-C_1^{(b)\mu}\right)\f{p_b^\nu}{m_b} \xi_a-\left(C_1^{(a)\nu}-C_1^{(b)\nu}\right)\f{p_b^\mu}{m_b} \xi_a }\Bigg] +\OO(\lambda^{-3})\, ,
\een
where $\xi_a$ is defined as $\xi_a\equiv \frac{\ln\tau_a}{\tau_a}$.

We shall now substitute this into \refb{eq:trajectory_newton} and compare the
two sides. For this we need to compute the first and second derivative of $X_a^\mu$
using the ansatz
\refb{eansatz}. We get,
\ben \label{e2.13}
\f{dX_a^\mu(\tau_a)}{d\tau_a}&=& \f{p_a^\mu}{m_a} +\OO(\lambda^{-1})\, ,\\ \label{e2.14}
\f{d^2X_a^\mu(\tau_a)}{d\tau_a^2}&=&-C_1^{(a)\mu}\f{1}{\tau_a^2}-\sum_{r=1}^\infty r(r+1)\ C_{r+1}^{(a)\mu} \f{(\ln\tau_a)^r}{\tau_a^{r+2}} +\OO(\lambda^{-3})\, .
\een
We can now provide a justification of why
self-force effects are not important in  \refb{e2.14}. For this note
that each term on the right hand side of this equation scales as $\lambda^{-2}$. In
contrast, the self-force gives terms proportional to the time derivative of the
acceleration which scales as $\lambda^{-3}$ or square of the acceleration which
scales as $\lambda^{-4}$. Thus they can be ignored. A similar justification can be
provided for ignoring the effects of multipole moments of the outgoing
particles. The multipole moments 
have two effects on our analysis. First of all for a given background electromagnetic
field, these produce additional terms to the right hand side of
\refb{e2.14} that involve derivatives of
the field strength. These are suppressed by additional powers of $\lambda^{-1}$.
The second effect is that the presence of the multipole moments can produce additional
contribution to the expression \refb{eq:F_b} for the
electromagnetic field that falls off faster than the inverse square
law at large distances. These are also suppressed by 
additional powers of $\lambda^{-1}$. Thus
we can ignore their effects.

Substituting \eqref{e2.11}, \refb{e2.13} and
\refb{e2.14} into \refb{eq:trajectory_newton}, we get,
\ben\label{e113rep}
&&\left( -C_1^{(a)\mu}-\sum_{r=1}^\infty r(r+1)\ C_{r+1}^{(a)\mu}\ \xi_a^r\right)\non\\
&\simeq& 
-\sum_{b\neq a}\f{q_aq_b}{4\pi} \f{p_b^2}{\left[(p_a.p_b)^2-p_a^2 p_b^2\right]^\f{3}{2}}\non\\
&&\times \left[\left(p_a^2 p_b^\mu -p_a.p_b p_a^\mu\right) +m_a p_a.p_b\left(C_1^{(b)\mu}-C_1^{(a)\mu}\right)\xi_a -m_a \left(C_1^{(b)}-C_1^{(a)}\right).p_a p_b^\mu\, \xi_a  \right]\non\\
&&\times \left[ 1+2m_a  \f{ p_b^2p_a.\left(C_1^{(b)}-C_1^{(a)}\right) -p_a.p_b p_b.\left(C_1^{(b)}-C_1^{(a)}\right) }{(p_a.p_b)^2 -p_a^2 p_b^2}\, \xi_a \right.\non\\
&&\hskip .5in \left.+m_a^2 \f{  \left(p_b.\left(C_1^{(b)}-C_1^{(a)}\right)\right)^2-p_b^2 \left(C_1^{(b)}-C_1^{(a)}\right)^2 }{(p_a.p_b)^2 -p_a^2 p_b^2}\, \xi_a^2\right]^{-\f{3}{2}}.
\een 
These equations can be solved iteratively to give:\footnote{In the RHS 
of \eqref{e113rep}, we used the following series expansion identity
\be
\left(1+A\xi_a +B\xi_a^2\right)^{-\f{3}{2}}=\sum_{k=0}^\infty \sum_{s=0}^k (-1)^k \f{(2k+1)!}{2^{2k}k! s!(k-s)!}A^s B^{k-s}\xi_a^{2k-s}.
\ee}
\be\label{eC1exp}
C_1^{(a)\mu}=\sum_{b\neq a} \f{q_a q_b}{4\pi} \f{p_b^2}{\left[ (p_a.p_b)^2-p_a^2 p_b^2\right]^{\f{3}{2}}}\left(p_a^2 p_b^\mu -p_a.p_b p_a^\mu\right),
\ee
\ben \label{eCrexp}
&&C_{n+1}^{(a)\mu}\non\\
&&=\f{m_a^n}{2^nn(n+1)}\sum_{b\neq a}\f{q_a q_b}{4\pi} \f{p_b^2}{\left[(p_a.p_b)^2-p_a^2 p_b^2\right]^\f{3}{2}} 
\Bigg[ \left(p_a^2 p_b^\mu -p_a.p_b p_a^\mu\right)\non\\
&&\times \sum_{r=\f{n}{2}\ 
{\rm for}\, n \, {\rm even}\atop
 r=\f{n+1}{2}\ {\rm for} \, n\, {\rm odd}}^n \f{(-1)^r\ (2r+1)!}{ r! (2r-n)! (n-r)!}  \f{1}{\left((p_a.p_b)^2-p_a^2 p_b^2\right)^r}\non\\
 &&\times \left(  p_b^2\, p_a.C_1^{(b)}
+p_a.p_b p_b.C_1^{(a)}\right)^{2r-n}   \left(  \left(p_b.C_1^{(a)}\right)^2-p_b^2 \left(C_1^{(b)}-C_1^{(a)}\right)^2 \right)^{n-r}\non\\
&& +\ 2\left( p_a.p_b\left(C_1^{(b)\mu}-C_1^{(a)\mu}\right) 
- p_a.\, C_1^{(b)}\,  p_b^\mu  \right) \non\\ &&
\times \ \sum_{r=\f{n}{2}\ {\rm for} \, n \, {\rm even}\atop r=\f{n-1}{2}\ {\rm for}\,
n \, {\rm odd}}^{n-1} \f{(-1)^r\ (2r+1)!}{ r! (2r+1-n)! (n-r-1)!} \non\\
&&\times \f{1}{\left((p_a.p_b)^2-p_a^2 p_b^2\right)^r}
\left(   p_b^2\, p_a.C_1^{(b)} 
+p_a.p_b p_b.C_1^{(a)} \right)^{2r+1-n}\non\\
&&\times  \left(  \left(p_b.C_1^{(a)}\right)^2-p_b^2 \left(C_1^{(b)}-C_1^{(a)}\right)^2\right)^{n-r-1}\Bigg]\, ,  \qquad \hbox{for $n\ge 1$}\, .
\een
In arriving at \refb{eCrexp} we have used the relation $p_a.C_1^{(a)}=0$ that follows from
\refb{eC1exp}.
These reproduce \refb{eC1expint}, \refb{eCrexpint}. The analysis of the trajectory for early time is similar. One difference is that
the factors of $\ln\tau_a$ are replaced by $\ln|\tau_a|$. 
The other difference is that due to the absolute value sign in the denominator of
\refb{eq:F_b}, one of the factors of $\tau_a$ in \refb{e2.11} is replaced by 
$|\tau_a|=-\tau_a$, leading to a change in sign of the right hand sides of
\refb{e113rep}. This can be removed by a redefinition $C_r^{(a)\mu}\to
(-1)^r C_r^{(a)\mu}$ and $\xi_a\to -\xi_a$. The net effect is that in the final expressions
\refb{eC1exp} and \refb{eCrexp} of $C_r^{(a)\mu}$ we have an extra factor of
$(-1)^r$, besides replacement of the outgoing momenta and charges by incoming
momenta $p_a^{\prime\mu}$ and charges $q'_a$.

\paragraph{Trajectory for two-body scattering:}
When there are only two particles in the final state, the analysis simplifies, since it
follows from \refb{eC1exp} that 
\ben
&& 1+2m_a  \f{ p_b^2p_a.\left(C_1^{(b)}-C_1^{(a)}\right) -p_a.p_b p_b.\left(C_1^{(b)}-C_1^{(a)}\right) }{(p_a.p_b)^2 -p_a^2 p_b^2}\, \xi_a \non\\
&&\hskip .5in +\ m_a^2 \f{  \left(p_b.\left(C_1^{(b)}-C_1^{(a)}\right)\right)^2-p_b^2 \left(C_1^{(b)}-C_1^{(a)}\right)^2 }{(p_a.p_b)^2 -p_a^2 p_b^2}\, \xi_a^2
\non \\
&=& \left[1+m_a  \f{ p_b^2\, p_a.\left(C_1^{(b)}-C_1^{(a)}\right) -p_a.p_b \, p_b.\left(C_1^{(b)}-C_1^{(a)}\right) }{(p_a.p_b)^2 -p_a^2 p_b^2}\, \xi_a\right]^2\non\\
&=&\left(1+\f{\xi_a}{m_a} \f{q_a q_b}{4\pi} \f{p_a^2 p_b^2 p_a.(p_a+p_b)}{\left[(p_a.p_b)^2-p_a^2 p_b^2\right]^\f{3}{2}}\right)^2\, .
\een
This simplifies the expansion of the right hand side of \refb{e113rep} in powers of
$\xi_a$ and we get,
\be \label{e1.11new}
C_{r+1}^{(a)\mu}= \f{m_a^r}{r}\,     
\left( \f{q_a q_b}{4\pi} \f{ p_b^2 }{\left[(p_a.p_b)^2-p_a^2 p_b^2\right]^\f{3}{2}}
\right)^{r+1}\ \left[p_a.(p_a+p_b)\right]^r\,  \left(p_a^2 p_b^\mu -p_a.p_b p_a^\mu\right),
\qquad b\ne a.
\ee
This is the result quoted in \refb{e1.11} and a resummed trajectory expression follows from this is quoted in \eqref{eq:trajectory_resummed}. 

\paragraph{Trajectory in probe approximation:} Here, we analyze the spatial component of the late-time multi-particle waveform expression \eqref{eq:log_n_waveform} in the probe limit. We consider a scattering event in which $(N-1)$ probe particles, characterized by charges $\lbrace q_a\rbrace$, masses $\lbrace m_a\rbrace$, and outgoing momenta $\lbrace p_a\rbrace$ for $a=2,3,\cdots ,N$, scatter in the presence of a heavy scatterer with mass $M$ and charge $Q$ at rest. The probe limit is defined as $M \gg m_a$, $Q \gg q_a$, and $\frac{Q}{M} \ll \frac{q_a}{m_a}$ for $a=2,3,\cdots ,N$. The extra condition $\frac{Q}{M} \ll \frac{q_a}{m_a}$ is needed to ensure that the heavy scatterer does not move under the electromagnetic force exerted by light probes and that the
electromagnetic radiation from it can be ignored. The momenta of the 
heavy scatterer and probe objects are parametrized as follows:
\ben
P=\left(M,\vec{0}\right),\hspace{2cm} p_a=\f{m_a}{\sqrt{1-\vec{\beta}_a^2}}\left(1,\vec{\beta}_a\right),\hspace{1cm}\forall a=2,3,\cdots N.
\een
In this probe limit, the trajectory coefficients in \eqref{eq:trajectory_patternb} for the probe particles will receive contributions solely from the long-range electromagnetic force exerted by the heavy scatterer. The leading trajectory correction $C_1^{(a)}$ given in \eqref{eC1exp} simplifies to
\ben\label{eq:C1_probe}
C_1^{(a)}&\simeq &-\f{q_aQ}{4\pi} \f{\sqrt{1-\beta_a^2}}{m_a\beta_a}\left(1,\, \f{\vec{\beta}_a}{\beta_a^2}\right),\hspace{1cm}\forall a=2,3,\cdots N.
\een
After substituting the above expression in \eqref{e113rep}, in the probe limit it also simplifies to 
\ben
&&\left( -C_1^{(a)}-\sum_{r=1}^\infty r(r+1)\ C_{r+1}^{(a)}\ \xi_a^r\right)\non\\
&\simeq & \f{q_aQ}{4\pi} \f{\sqrt{1-\beta_a^2}}{m_a\beta_a}\left(1,\f{\vec{\beta}_a}{\beta_a^2}\right)\left(1-\f{q_a Q}{4\pi} \f{1-\beta_a^2}{\beta_a^3}\f{\xi_a}{m_a}\right)^{-2}.
\een 
Performing a series expansion in $\xi_a$ and comparing the coefficient of $\xi_a^r$ from both sides, we get
\ben\label{eq:Cr+1_probe}
C_{r+1}^{(a)}= -\f{1}{r}\f{q_aQ}{4\pi} \f{\sqrt{1-\beta_a^2}}{m_a\beta_a}\left(\f{q_a Q}{4\pi} \f{1-\beta_a^2}{m_a\beta_a^3}\right)^r\times \left(1,\, \f{\vec{\beta}_a}{\beta_a^2}\right). 
\een
The above trajectory coefficient matches \eqref{e1.11new} after simplifying \eqref{e1.11new} by considering particle-$b$ as the heavy scatterer. Hence, an analogous resummed trajectory can be provided in the probe limit, as quoted in \eqref{eq:trajectory_resummed} for two-particle scattering.

\sectiono{Electromagnetic wave-form} \label{ewaveform}

Once the trajectories of the outgoing particles have been determined, we can use
\refb{eq:Ab_mu_int_rep} to find the electromagnetic field at the detector after summing over $b$. The analysis
follows the same procedure that led to the expression \refb{e2.11} for the field strength,
except that the approximation scheme that we shall use will have to
be somewhat different.
If the detector is placed at a distance $R$ from the region where the scattering takes
place, then we first expand $A_\mu$ in powers of $1/R$ and keep only terms of order
$1/R$, and then expand the coefficient of this term in inverse powers of the retarded
time $u$.

Let us parametrize the location $x$ of the detector as $x^\mu=(u+R,R\hat{n})
 = (u,\vec 0)+R\,\mathbf{n}$. 
For large $R$ and fixed $u$, the delta function constraint in \eqref{eq:Ab_mu_int_rep} implies the following relation:
\ben\label{e3.1}
&&\left(x-X_a(\tau_a)\right)^2=0\hspace{1cm}\Rightarrow\hspace{1cm} -2uR -2R\ \mathbf{n}.X_a(\tau_a)+\OO(R^0)=0.
\een
Using \refb{eansatz},
the leading $R$ coefficient of the above equation gives
\ben\label{e3.2}
-u&=& \f{\mathbf{n}.p_a}{m_a} \tau_a +\mathbf{n}.C_1^{(a)} \ln\tau_a -\sum_{r=1}^\infty \mathbf{n}.C_{r+1}^{(a)} \left(\f{\ln\tau_a}{\tau_a}\right)^{r}\, +\OO(\lambda^{-1}).
\een
Note that on the right hand side the first two terms are of order $\lambda$ and the other
terms are of order $\lambda^0$. Hence we keep only the first two terms, and invert
the equation to find the solution 
\ben\label{e115rep}
\tau_a=\tau^{\rm sol}_a,\hspace{1cm}
&&\tau_a^{\rm sol} \simeq -{u\, m_a\over \bn.p_a}  \left( 1+ {\bn.C_1^{(a)}}\, w\right),
\quad w\equiv {\ln u\over u},\non\\
 \Rightarrow && 
\xi_a\equiv {\ln\tau_a^{\rm sol}\over \tau_a^{sol} }
\simeq -{\bn.p_a\over m_a} \, w\, (1 + \bn.C_1^{(a)}\, w)^{-1}\,  .
\een 
Hence for large value of $R$ and late retarded time $u$ 
the relevant part of the delta function constraint in \eqref{eq:Ab_mu_int_rep} becomes
\ben
\delta_+\left(-\left(x-X_a(\tau_a)\right)^2\right)&\simeq &\f{1}{2R} \f{\delta\left(\tau_a-\tau_a^{sol}\right)}{\Big{|} \mathbf{n}.\f{dX_a(\tau_a)}{d\tau_a}\Big{|}_{\tau_a=\tau_a^{sol}}}.
\een
We also have
\be\label{e3.5}
{dX^\mu_a\over d\tau_a} \simeq \f{p_a^\mu}{m_a}  +C_1^{(a)\mu} \f{1}{\tau_a} +\sum\limits_{r=1}^\infty r C_{r+1}^{(a)\mu} \f{(\ln\tau_a)^r}{\tau_a^{r+1}}\, .
\ee
Note that the first term on the right hand side is of order $\lambda^0$ while the other
terms are of order $\lambda^{-1}$. Thus we might be tempted to keep only the first
term. However we shall be computing terms in $A_\mu$ to first subleading order in
the expansion in $\lambda^{-1}$, and for this reason we have to keep all the terms.
Using the above results in \eqref{eq:Ab_mu_int_rep}, the late time radiative mode of electromagnetic wave-form becomes
\be\label{e66}
A^{\mu}(u,R\hat{n})\simeq - \f{1}{4\pi R}\sum_{a} \f{m_a q_a}{\mathbf{n}.p_a}  \int_0^\infty d\tau_a\,  
\delta\left(\tau_a-\tau_a^{sol}\right) \f{\left( \f{p_a^\mu}{m_a}  +C_1^{(a)\mu} \f{1}{\tau_a} +\sum\limits_{r=1}^\infty r\, C_{r+1}^{(a)\mu} \f{(\ln\tau_a)^r}{\tau_a^{r+1}}\right)}{1+\f{m_a}{\mathbf{n}.p_a} \f{\mathbf{n}.C_1^{(a)}}{\tau_a} +\f{m_a}{\mathbf{n}.p_a} \sum\limits_{r=1}^\infty r\, \mathbf{n}.C_{r+1}^{(a)}\f{(\ln\tau_a)^r}{\tau_a^{r+1}}  }\, ,
\ee
where we used the fact that $\mathbf{n}.p_a$ is negative. Performing the integral over $\tau_a$, and keeping terms up to order 
$\lambda^{-1}$, we get
\ben\label{e117arep}
A^{\mu}(u,R\hat{n})&\simeq &- \f{1}{4\pi R}\sum_{a} \f{q_ap_a^\mu}{\mathbf{n}.p_a}  
 \nonumber \\  
&&\hskip -.2in  - \f{1}{4\pi R}
\sum_{a} {1\over \tau^{sol}_a}\f{m_a q_a}{\mathbf{n}.p_a}  
\left( C_1^{(a)\mu} - 
\f{p_a^\mu}{\mathbf{n}.p_a} \mathbf{n}.C_1^{(a)} +
\sum_{r=1}^\infty r C_{r+1}^{(a)\mu} \xi_a^r -
\f{p_a^\mu}
{\mathbf{n}.p_a}  \sum\limits_{r=1}^\infty r \mathbf{n}.C_{r+1}^{(a)}
\xi_a^r \right)\,. \non\\
\een
An expansion in $w$ of the above expression after using \eqref{e115rep} generates the $u^{-n}(\ln u)^{n-1}$ coefficients to the electromagnetic wave-form:
\ben\label{eq:log_n_waveformrep}
&&A^{\mu}(u,R\hat{n})\non\\
&\simeq  &- \f{1}{4\pi R}\sum_{a} \f{q_ap_a^\mu}{\mathbf{n}.p_a} \non\\
&&+  \f{1}{4\pi R}\sum_{n=1}^\infty (-1)^n \f{(\ln u)^{n-1}}{u^n}\sum_{a}\f{q_a}{p_a.\mathbf{n}}\Bigg[ \left( \mathbf{n}.C_1^{(a)}\right)^{n-1} \left( \mathbf{n}.C_1^{(a)} p_a^\mu -\mathbf{n}.p_a  C_1^{(a)\mu} \right) \non\\
&&+\sum_{r=1}^{n-1} \f{(n-1)!}{(r-1)! (n-r-1)!} \left(\mathbf{n}.C_{r+1}^{(a)} p_a^\mu-\mathbf{n}.p_a C_{r+1}^{(a)\mu} \right) \left(\f{\mathbf{n}.p_a}{m_a}\right)^r  \left( \mathbf{n}.C_1^{(a)}\right)^{n-r-1}\Bigg]\, .
\een
The result for large negative $u$ is obtained from \eqref{eq:log_n_waveformrep} by
replacing $q_a$, $p_a^\mu$ by the incoming charges and momenta $q'_a$, $p_a^{\prime
\mu}$, $\ln u$ by $\ln |u|$ and including an extra factor of $(-1)^r$ in the expression
for $C_r^{(a)\mu}$ due to the observations made below \eqref{e1.11new}.

Note that the first term on the right hand side 
of \refb{eq:log_n_waveformrep} is of order $\lambda^0$ while the
rest of the terms are of order $\lambda^{-1}$. This explains the reason 
for keeping
the first subleading terms in the expression for $dX^\mu_a/d\tau_a$ in \refb{e3.5}.
However, we kept only the leading term in the expression for $X^\mu_a$ in
\refb{e3.2} while finding the solution for $\tau_a$ in \refb{e115rep}. This can be justified
by noting that the leading term in the integrand of \refb{e66} is $\tau_a$ independent.
Thus a subleading correction to $\tau_a^{\rm sol}$ in \refb{e115rep} 
will not give any subleading correction
to \refb{eq:log_n_waveformrep}.

In this convention $A_\mu$ in \eqref{eq:log_n_waveformrep} 
does not vanish as $u\to\pm\infty$. It is more
convenient to express $A_\mu$ by adding a constant to it so that it vanishes
as $u\to -\infty$, which we refer to observer's frame. In that case the same constant will be added to $A_\mu$ in the
large positive $u$ limit and the constant term (first line of RHS) in
\refb{eq:log_n_waveformrep} will be replaced by\footnote{Due to the dependence on
$R$ and $\mathbf{n}$, the term in the first line of \refb{eq:log_n_waveformrep} depends
on the spatial coordinates and hence cannot be removed altogether by a
gauge transformation. However these spatial derivatives are of order $R^{-2}$ for
large $R$, and hence as long as we ignore terms of order $R^{-2}$, we can remove these
terms in the local neighborhood of the observer by a gauge transformation.}
\be \label{e3.8}
- \f{1}{4\pi R}\sum_{a} \f{q_ap_a^\mu}{\mathbf{n}.p_a} +
 \f{1}{4\pi R}\sum_{a} \f{q'_ap_a^{\prime\mu}}{\mathbf{n}.p'_a} .
\ee
This is known as the electromagnetic memory term. This is the procedure followed when writing the constant term in \eqref{eq:log_n_waveform}.

\paragraph{Wave-form for two-body scattering:}
For $2\to 2$ scattering, we can use \refb{e1.11new}, \refb{eq:log_n_waveformrep} 
and
\refb{e3.8} to get,
for large positive $u$,
\ben\label{eq:log_n_waveform_2particle}
&&A^{\mu}(u,R\, \hat{n})\non\\
&\simeq&- \f{1}{4\pi R}\sum_{a} \f{q_ap_a^\mu}{\mathbf{n}.p_a}  +
 \f{1}{4\pi R}\sum_{a} \f{q'_ap_a^{\prime\mu}}{\mathbf{n}.p'_a}\non\\
&& + \f{1}{4\pi R}  \sum_{n=1}^\infty
(-1)^{n}\f{(\ln u)^{n-1}}{u^n}\sum_{a,b\ne a}  
\left(\sigma_{ab}^{out}\right)^{n}\f{q_a}{p_a.\mathbf{n}}\left(\mathbf{n}.p_b p_a^\mu -\mathbf{n}.p_a p_b^\mu\right) \left( (p_a+p_b).\mathbf{n}\right)^{n-1}\, ,
\een
where $\sigma_{ab}^{out}$ has been defined in \refb{edefsigmaabout}. This is quoted in \eqref{eq:log_n_waveform_2particleint}, and after performing the sum over $n$ it reproduces \eqref{eq:resummed_u_2particle_late}. The results for large negative $u$ are obtained by dropping
the constant term and replacing $u^n$ by $|u|^n$, $\ln u$ by $\ln|u|$ and  the outgoing charges and momenta by the incoming charges and momenta $\{q'_a\}$
and $\{p'_a\}$. Following this prescription a resummed early retarded time wave-form is presented in \eqref{eq:resummed_u_2particle_early}.

We now use the inverse Fourier transform of the following results derived in \cite{2008.04376}:
\ben
\int_{-\infty}^{\infty}\f{d\omega}{2\pi}\ e^{-i\omega u}\ \ \omega^{n-1}\Big{\lbrace}\ln(\omega +i\epsilon)\Big{\rbrace}^{n} &\simeq &\ \left\{ \begin{array}{ll} - n!\ i^{n-1}\ \f{(\ln |u|)^{n-1}}{u^{n}} \hspace{1cm}\hbox{for\ }\ u\rightarrow +\infty\\
\\
0\hspace{4cm}\hbox{for\ }\ u\rightarrow -\infty
\end{array}\right. ,\vspace{1cm}
\een
\ben
\int_{-\infty}^{\infty}\f{d\omega}{2\pi}\ e^{-i\omega u}\  \omega^{n-1}\Big{\lbrace}\ln(\omega -i\epsilon)\Big{\rbrace}^{n} &\simeq &\ \left\{ \begin{array}{ll} 0\hspace{4cm}\hbox{for\ }\  u\rightarrow +\infty\\
\\
+ n!\ i^{n-1}\ \f{(\ln |u|)^{n-1}}{u^{n}} \hspace{1cm}\hbox{for\ }\  u\rightarrow -\infty 
\end{array}\right. ,
\een
to write the frequency space wave-form:
\ben\label{eq:log_n_waveform_2particle_frequencyrep}
&& \wt A^\mu (\omega, \vec x)\non\\
&\simeq&-\f{i}{4\pi R} \f{1}{\omega  + i\eps }
\sum_{a=1,2}\left(\f{q_ap_a^\mu}{p_a.\mathbf{n}}-\f{q'_ap_a^{\prime\mu}}{p'_a.\mathbf{n}}\right)\non\\
&&- \f{i}{4\pi R}\sum_{n=1}^\infty \f{1}{n!}
\omega^{n-1}\lbrace \ln(\omega+i\epsilon)\rbrace^n \sum_{a,b\atop b\ne a}  
\left(i\sigma_{ab}^{out}\right)^{n}\f{q_a}{p_a.\mathbf{n}}\left(\mathbf{n}.p_b p_a^\mu -\mathbf{n}.p_a p_b^\mu\right) \left( (p_a+p_b).\mathbf{n}\right)^{n-1} \non\\ &&+
 \f{i}{4\pi R}\sum_{n=1}^\infty \f{1}{n!}
\omega^{n-1}\lbrace \ln(\omega-i\epsilon)\rbrace^n \sum_{a,b\atop b\ne a}  
\left(-i\sigma_{ab}^{in}\right)^{n}\f{q'_a}{p'_a.\mathbf{n}}\left(\mathbf{n}.p'_b 
p_a^{\prime \mu} 
-\mathbf{n}.p'_a p_b^{\prime\mu}\right) \left( (p'_a+p'_b).\mathbf{n}\right)^{n-1} \, .\non\\
\een
This is quoted in \eqref{eq:log_n_waveform_2particle_frequency}, and after performing the sum over $n$ it reproduces \eqref{eq:resummed_em_waveform}.

\paragraph{Wave-form in probe approximation:}Here, we analyze the spatial component of the late-time multi-particle waveform expression \eqref{eq:log_n_waveform} in the probe limit. After substituting the trajectory coefficients \eqref{eq:C1_probe} and \eqref{eq:Cr+1_probe} in \eqref{eq:log_n_waveform}, we get 
\ben
&&A^{i}(u,R\, \hat{n})\non\\
&\simeq &\f{1}{4\pi R} \sum_{a=2}^N q_a \left(\f{\beta_a^i}{1-\hat{n}.\vec{\beta}_a}-\f{\beta_a^{\prime i}}{1-\hat{n}.\vec{\beta}^\prime_a}\right) \non\\
&&-\f{1}{4\pi R}\sum_{n=1}^\infty \f{(\ln u)^{n-1}}{u^n}\sum_{a=2}^N   \left(\f{Qq_a}{4\pi} \f{(1-\beta_a^2)^\f{3}{2}}{m_a\beta_a^3}\right)^{n}  \times \f{q_a\beta_a^i }{1-\hat{n}.\vec{\beta}_a},
\een
where $\lbrace \vec{\beta}^\prime_a\rbrace$ denote the velocities of ingoing probe particles. $A^0(u, R \hat{n})$ can be found from this using the relation 
$\mathbf{n}.A(u,R \hat n)=0$.
The $1/u$ coefficient of the above expression agrees with the result of \cite{1804.09193}. Here, we note that the above waveform expression is essentially the sum over $a=2,3,\cdots ,N$ of the late-time waveform for $2 \rightarrow 2$ scattering, as given in \eqref{eq:log_n_waveform_2particle}, once we express it by 
treating particle $a$ as the probe and particle $b$ as the heavy scatterer. 
This feature resembles the result for the low-frequency gravitational waveform in the probe approximation presented in \cite{2408.07329}.

\sectiono{Discussion} 

Our results \refb{eq:log_n_waveform_2particle_frequencyrep} and \eqref{eq:resummed_em_waveform_simplified} closely
resemble the conjectured result of \cite{2407.04128} for low frequency gravitational
wave emitted during a $2\to 2$ scattering process.
However we have also seen that when the number of incoming or outgoing 
particles exceeds two, the formula is more complicated due to the
more complicated expressions for
$C_{r+1}^{(a)\mu}$'s as given in \refb{eCrexp}.
This suggests that the conjecture of \cite{2407.04128} 
is likely to be true for scatterings when neither the
number of incoming particles nor the number of outgoing particles exceeds two,
but is likely to be modified for more external particles. It will clearly be of interest to
generalize the analysis of this paper to gravitational wave emission during a scattering
process, and we hope to return to this problem in an upcoming paper \cite{all_order_soft_gr}.

The simplicity of the resummed leading-log wave-form result in \eqref{eq:resummed_em_waveform_simplified} suggests that, in the analysis of the all-loop QED S-matrix for $2 \rightarrow 2$ scattering, one should be able to derive an analogous leading-log resummed soft photon theorem. Following \cite{1801.07719,1808.03288}, we conjecture that
the quantum soft factor for a single soft photon emission, 
which appears as the ratio of the QED S-matrix with one external soft photon 
to the S-matrix without the soft photon, can be expressed as:
\be
\mathbb{S}_{em}=i(4\pi R)\varepsilon_{\mu}(k)\widetilde{A}_{Q}^\mu(\omega, R\hat{n}), \hspace{1cm} k\equiv \omega (1,\hat{n}),
\ee
where $\widetilde{A}_{Q}^\mu$ is related to the frequency-space 
wave-form $\widetilde{A}^\mu$ given in \refb{eFT}
as follows.
First, since $\omega$ represents the external photon energy, we do not
use any $i\eps$ prescription for $\omega$ and
$\ln(\omega \pm i\epsilon)$ is replaced by
$\ln \omega$. Also, 
the `trajectory coefficients' $C_r^{(a)\mu}$ need to be evaluated using the Feynman propagator instead of the retarded propagator used here. 
Due to the use of Feynman propagator $C_r^{(a)\mu}$ receives contribution from both ingoing and outgoing particles. By extending the one-loop and two-loop analysis of \cite{1808.03288,2308.16807,2008.04376} to arbitrary loop orders, it 
should be
possible to explicitly derive the leading-log soft photon theorem and test the above proposal.

One could also explore various generalizations of the results of this paper. For example,
the order $u^{-2} $ gravitational waveform in the leading post-Minkowskian expansion under purely gravitational interaction was derived in \cite{2106.10741}, and it depends on the momenta, spins, and impact parameters of the scattering process. This result for two particle scattering has been tested in \cite{2310.05832,2309.14925}. The analog
of this result for the electromagnetic scattering will be a formula for the subleading 
(in $\lambda$) terms in the wave-form that depends on the impact parameters and the
multipole moments of the external particles.
Recently, for two-particle scattering with impact parameter $b$,
an order $(\omega b)^n\ln(\omega b)$ contribution to the 
gravitational waveform for arbitrary $n$ was derived in \cite{2409.12898} in the
limit $\omega\to 0$, $b\to\infty$ with $\omega b$ fixed. This is a different limit,
and it may be interesting to explore the electromagnetic analog of this.

\bigskip

\noindent {\bf Acknowledgement:} 
We would like to thank Samim Akhtar, Carlo Heissenberg, R.~Loganayagam,  Ana-Maria Raclariu,
Suvrat Raju, Omkar Shetye and Kaustubh Singhi for useful discussions. We also thank
Carlo Heissenberg for many useful comments and suggestions on the
earlier version of the draft. 
D.K. and B.K. were supported by the 
Department of Atomic Energy, Government of India, under project no. RTI4001. B.S. was supported by STFC grant number ST/X000753/1. A.S. was supported by the ICTS-Infosys Madhava 
Chair Professorship
and the Department of Atomic Energy, Government of India, under project no. RTI4001.

\end{document}